\documentclass[a4paper,12pt]{article}

\usepackage{jheppub}

\usepackage{cancel}
\usepackage{graphicx}
\usepackage{color}
\usepackage{dsfont}
\usepackage{overpic}
\usepackage{epic}
\usepackage{epstopdf}

\usepackage{amsmath,amssymb}

\newlength{\nseparation}
\newcommand{\bea}{\begin{eqnarray}}
\newcommand{\eea}{\end{eqnarray}}

\newcommand{\pslash}{p\hspace{-0.44em}/\hspace{0.06em}}

\newcommand{\eq}[1]{Eq.~(\ref{#1})}

\newcommand{\Br}{\text{Br}}

\newcommand{\matrixx}[1]{\begin{pmatrix} #1 \end{pmatrix}}

\preprint{PSI-PR-19-02, ZU-TH 10/19}

\title{\begin{boldmath}$b\to s\ell^+\ell^-$\end{boldmath} Transitions in Two-Higgs-Doublet Models}

\author[a,b]{Andreas Crivellin,}
\author[a,b]{Dario M\"uller}
\author[c]{Christoph Wiegand}

\affiliation[a]{Paul Scherrer Institut, CH--5232 Villigen PSI, Switzerland}
\affiliation[b]{Physik-Institut, Universit\"at Z\"urich, Winterthurerstrasse 190, CH-8057 Z\"urich, Switzerland}
\affiliation[c]{Albert Einstein Center for Fundamental Physics, Institute
	for Theoretical Physics, University of Bern, CH-3012 Bern,
	Switzerland}

\emailAdd{andreas.crivellin@psi.ch}
\emailAdd{dario.mueller@psi.ch}
\emailAdd{wiegand@itp.unibe.ch}

\abstract{In this article we study $b\to s\mu^+\mu^-$ transitions and possible correlations with the anomalous magnetic moment of the muon ($a_\mu$) within two-Higgs-doublet models with generic Yukawa couplings, including the possibility of right-handed neutrinos. We perform the matching on the relevant effective Hamiltonian and calculate the leading one-loop effects for $b\to s\ell\ell^{(\prime)}$, $b\to s\gamma$, $\Delta B=\Delta S=2$, $b\to s\nu\bar\nu$ and $\ell\to\ell^\prime\gamma$ transitions in a general $R_\xi$ gauge. Concerning the phenomenology, we find that an explanation of the hints for new physics in $b\to s\mu^+\mu^-$ data is possible once right-handed neutrinos are included. If lepton flavour violating couplings are allowed, one can account for the discrepancy in $a_\mu$ as well. However, only a small portion of parameter space gives a good fit to $b\to s\mu^+\mu^-$ data and the current bound on $h\to\tau\mu$ requires the mixing between the neutral Higgses to be very small if one aims at an explanation of $a_\mu$.}

\begin{document}

\maketitle

\newpage
\section{Introduction}
\label{sec:intro}

Two-Higgs-doublet models (2HDMs)~\cite{Lee:1973iz} have been under intensive investigation for a long time (see e.g. Ref.~\cite{Gunion:1989we} for an introduction or Ref.~\cite{Branco:2011iw} for a review article). There are several reasons for this intense interest: First of all, 2HDMs are extremely simple extensions of the Standard Model (SM) obtained by adding a single scalar $SU(2)_L$ doublet to the SM particle content. Furthermore, motivation for 2HDMs comes from axion models \cite{Kim:1986ax} because a possible CP violating QCD-theta term  can be absorbed~\cite{Peccei:1977hh} if the Lagrangian possesses a global $U(1)$ symmetry. This is only possible if the SM is extended by at least one Higgs doublet. Also the baryon asymmetry of the universe can be generated within 2HDMs while the amount of CP violation in the SM alone is too small to achieve this~\cite{Trodden:1998qg}. Finally, the Minimal Supersymmetric Standard Model predicts the presence of a second Higgs doublet~\cite{Haber:1984rc}, due to the holomorphicity of the superpotential. The effective theory obtained after integrating out the superpartners of the SM particles (sfermions, gaugions and higgsinos) is a 2HDM (with the addition of higher dimensional operators involving two Higgs doublets~\cite{Crivellin:2016ihg}). 
\medskip

2HDMs possess three additional physical scalars with respect to the single Higgs boson of the SM; a neutral CP-even $H^0$, a CP-odd scalar  $A^0$ and a charged scalar $H^\pm$ (under the assumption of CP conservation). These new particles are not only interesting with respect to direct searches at the LHC~(see e.g. Ref.~\cite{Bhatia:2017ttp,Arbey:2017gmh,Basler:2017nzu,Haisch:2017gql,Jenniches:2018zlb,Chen:2018uim,Enberg:2018pye,Arhrib:2018ewj,Hanson:2018uhf} for recent reports). In addition, they give rise to important effects in low-energy precision flavour observables, providing a complementary window to physics beyond the SM. In this respect, decays of neutral mesons to charged lepton pairs (e.g. $B_{s(d)}\to\mu^+\mu^-$, $D\to\mu^+\mu^-$ and $K_L\to\mu^+\mu^-$) are very interesting because they are especially sensitive to scalar operators which possess enhanced matrix elements with respect to vector operators. For this reason, $B_s\to\mu^+\mu^-$ (which can be calculated more precisely than $D\to\mu^+\mu^-$ or $K_L\to\mu^+\mu^-$ and has a larger branching fraction than $B_d\to\mu^+\mu^-$) has been studied frequently in the context of 2HDMs. However, the focus was on models with natural flavour conservation (i.e. with a $\mathds{Z}_2$ symmetry in the Yukawa sector)~\cite{He:1988tf,Skiba:1992mg,Dai:1996vg,Huang:1998bf,Logan:2000iv,Cheng:2015yfu,Arnan:2017lxi}, alignment~\cite{Iltan:2000ck,Li:2014fea} or generic flavour violation in the down sector~\cite{Mahmoudi:2009zx,Buras:2010mh,Crivellin:2013wna,Crivellin:2017upt}. In all these setups, the dominant effect originates from scalar operators. The current measurement of $B_{s}\to\mu^+\mu^-$~\cite{Amhis:2016xyh} (by ATLAS, CMS and LHCb \cite{CMS:2014xfa,Aaij:2013aka,Chatrchyan:2013bka,Aaboud:2018mst})
\begin{equation}
{\rm Br}[B_s\to\mu^+\mu^-]_{\rm EXP}=(3.1\pm 0.7)\times 10^{-9}\,, \label{BsmumuEXP}
\end{equation}
agrees quite well with the SM prediction~\cite{Bobeth:2013uxa,Beneke:2017vpq}
\begin{equation}
{\rm Br}[B_s\to\mu^+\mu^-]_{\rm SM}=(3.57\pm 0.17)\times 10^{-9}\,. \label{BsmumuSM}
\end{equation}
This puts stringent constraints on 2HDMs with scalar operators contributing to $b\to s\mu^+\mu^-$ transitions. Furthermore, LHCb found significant hints for new physics in $b\to s\ell^+\ell^-$ data, showing a coherent pattern of deviations from the SM predictions with a significance of more than $4$--$5\,\sigma$~\cite{Capdevila:2017bsm,Alguero:2019ptt}\footnote{Including only $R(K)$ and $R(K^*)$, the significance is at the 4$\,\sigma$ level~\cite{Altmannshofer:2017yso,DAmico:2017mtc,Geng:2017svp,Ciuchini:2017mik,Hiller:2017bzc,Hurth:2017hxg,Ciuchini:2019usw}.}. However, in order to explain these anomalies, vector operators, in particular $O_9$, are necessary while an explanation of the anomalies with scalar operators alone is not possible. 
\medskip

Within 2HDMs, vector operators at the dimension 6 level can only be generated via loop effects. However, contributions to other loop-induced processes such as $b\to s\gamma$ (for which the SM prediction \cite{Misiak:2015xwa} is in very well agreement with the experimental average \cite{Amhis:2016xyh}), $b\to s\nu\bar{\nu}$, (where the experimental upper bound \cite{Grygier:2017tzo,Lees:2013kla} approaches the SM prediction \cite{Buras:2014fpa}) or $B_s-\bar B_s$ mixing~\cite{Amhis:2016xyh} unavoidably arise and their constraints must be taken into account. Therefore, an explanation of $b\to s\ell^+\ell^-$ data in the context of multi-Higgs-doublet models might require the introduction of right-handed neutrinos~\cite{Li:2018rax,Marzo:2019ldg}. Furthermore, any model with sizeable couplings to muons could potentially address the long-lasting discrepancy between experiment~\cite{Bennett:2006fi} and the SM prediction\footnote{The SM prediction of $a_\mu$ is currently re-evaluated in a community-wide effort prompted by upcoming improved measurements at Fermilab~\cite{Grange:2015fou} and J-PARC~\cite{Saito:2012zz} (see also~\cite{Gorringe:2015cma}). With electroweak~\cite{Czarnecki:1995wq,Czarnecki:1995sz,Gnendiger:2013pva} and QED~\cite{Aoyama:2017uqe} contributions under good control, recent advances in the evaluation of the hadronic part include: hadronic vacuum polarization~\cite{DellaMorte:2017dyu,Davier:2017zfy,Borsanyi:2017zdw,Blum:2018mom,Keshavarzi:2018mgv,Giusti:2018mdh,Colangelo:2018mtw}, hadronic light-by-light scattering~\cite{Gerardin:2016cqj,Blum:2016lnc,Colangelo:2017qdm,Colangelo:2017fiz,Blum:2017cer,Hoferichter:2018dmo}, and higher-order hadronic corrections~\cite{Kurz:2014wya,Colangelo:2014qya}.} 
\begin{equation}
\label{Delta_amu}
\Delta a_\mu=a_\mu^\text{EXP} - a_\mu^\text{SM}\sim 270(85)\times 10^{-11}\,,
\end{equation}
of $3$--$4\,\sigma$. For definiteness, and in order to be conservative, we choose a value at the lower end. In the case of lepton flavour violation, $a_\mu$ is intrinsically correlated to lepton flavour violating decays such as $\tau\to\mu\gamma$ whose bound must be taken into account. Furthermore, in 2HDMs also $h\to\tau\mu$ gives relevant bounds due to the mixing between the neutral CP-even Higgses.
\medskip

In this article we want to investigate $b\to s\mu^+\mu^-$ transitions within 2HDMs in the light of the corresponding hints for new physics and its correlations with other $b\to s$ transitions and $a_\mu$. For this purpose, we will consider a 2HDM with a CP conserving Higgs potential but with generic sources of flavour violation and the possible addition of right-handed neutrinos. After establishing our conventions in Sec.~\ref{Model}, we will use this setup to calculate the tree-level matching on the effective Hamiltonian governing $b\to s$ transitions and the leading one-loop effects in Sec.~\ref{bsll}. Section~\ref{calculation2} is devoted to the calculation of the matching on the $\Delta B=\Delta S=2$ Hamiltonian, to $a_\mu$, $h\to\tau\mu$ and $b\to s\nu\bar\nu$. In our phenomenological analysis in Sec.~\ref{phenomenology} we will address the question if the hints for new physics in $b\to s\mu^+\mu^-$ transitions can be explained within 2HDMs without violating the bounds from other processes, before we conclude in Sec.~\ref{conclusions}.
\medskip

\section{Model and Conventions}
\label{Model}

\begin{table}
	\centering
	\begin{tabular}{||c||c|c|c|c|c|c||}
		\hline\hline
		$\rm{Type}$& ${c_y^d}$&${c_y^u}$&${c_y^\ell }$&${c_{\tilde \varepsilon }^d}$&${c_{\tilde \varepsilon }^u}$&${c_{\tilde \varepsilon }^\ell }$\\
		\hline\hline
		I&$ {\cot \left( \beta  \right)}$&${\cot \left( \beta  \right)}$&${\cot \left( \beta  \right)}$&${ - \sin \left( \beta  \right)}$&${ - \sin \left( \beta  \right)}$&${ - \sin \left( \beta  \right)}$\\
		II&${ - \tan \left( \beta  \right)}$&${\cot \left( \beta  \right)}$&${ - \tan \left( \beta  \right)}$&${\cos \left( \beta  \right)}$&${ - \sin \left( \beta  \right)}$&${\cos \left( \beta  \right)}$\\
		X& ${\cot \left( \beta  \right)}$&${\cot \left( \beta  \right)}$&${ - \tan \left( \beta  \right)}$&${ - \sin \left( \beta  \right)}$&${ - \sin \left( \beta  \right)}$&${\cos \left( \beta  \right)}$\\
		Y&$ { - \tan \left( \beta  \right)}$&${\cot \left( \beta  \right)}$&${\cot \left( \beta  \right)}$&${\cos \left( \beta  \right)}$&${ - \sin \left( \beta  \right)}$&${ - \sin \left( \beta  \right)}$\\
		\hline\hline
	\end{tabular}
	\caption{Relations between the parameters $\varepsilon_{ij}^{F}$ of the Higgs basis and the new  parameters $\tilde{\varepsilon}^{F}_{ij}$ in one of the other four bases with $\varepsilon_{ij}^{F}=c_{y}^{F}y^f_{i}\delta_{ij}+\tilde{\varepsilon}^{F}_{ij}/c_{\tilde{\varepsilon}}^{F}$. The $\tilde{\varepsilon}^{F}_{ij}$ break the $\mathds{Z}_2$ symmetry of the four 2HDMs with natural flavour conservation and induce flavour changing neutral currents.
	}\label{tablebasis}
\end{table}

As outlined in the introduction, we supplement the SM by a second scalar doublet with the same hypercharge as the first one. For the calculation of flavour observables it is convenient to work in the Higgs basis~\cite{Georgi:1978ri,Lavoura:1994fv,Botella:1994cs} where only one Higgs doublet acquires a vacuum expectation value and therefore the generation of the fermions and gauge boson masses is separated from the couplings to fermions. Using the notation of Ref.~\cite{Davidson:2016utf}, we have
\begin{align}
\Phi_1 =\matrixx{ G^+ \\ \dfrac{v+H_1^0+i G^0}{\sqrt{2}}}, &&
\Phi_2 = \matrixx{H^+ \\ \dfrac{H_2^0 + i A^0}{\sqrt{2}}},
\label{Phidef}
\end{align}
with $v\simeq 246$ GeV. $G^{+}$ and $G^{0}$ are the Goldstone bosons and $A^0$ denotes the physical CP-odd scalar, assuming that CP is conserved in the Higgs potential. The CP-even mass eigenstates are 
\begin{align}
\begin{split}
h^0 &= H_1^0 \sin (\beta-\alpha) + H_2^0 \cos (\beta-\alpha)\,,\\
H^0 &= H_1^0 \cos (\beta-\alpha) - H_2^0 \sin (\beta-\alpha)\,,
\end{split}
\label{baRot}
\end{align}
where we defined the mixing angle as $\beta-\alpha$ for easier comparison with the well-known type-I/II/X/Y 2HDMs. In the following, we will abbreviate $s_{\beta\alpha}\equiv \sin (\beta-\alpha)$ and $c_{\beta\alpha}\equiv \cos (\beta-\alpha)$ and assume that $h^0$ is the SM-like Higgs boson with a mass of around 125 GeV. We require $c_{\beta\alpha}$ to be small (at most $\mathcal{O}(0.1)$) such that its properties are compatible with experiments~\cite{Sirunyan:2018koj,ATLAS:2018doi}. With these conventions the couplings of the scalar bosons to fermions are given by
\begin{align}
\begin{split}
{L_Y} &= -\!\!\! \sum\limits_{F = u,d,\ell ,\nu } \bigg[ 	{{\bar F}_f}\Big(\, {\dfrac{{m_f^F}}{v}{\delta _{fi}} c_{\beta\alpha} \!- \!\left( {\varepsilon _{fi}^F{P_R} + \varepsilon _{if}^{F*}{P_L}} \right)\!s_{\beta\alpha}} \Big)\!{F_i}{H^0}\\
	&\phantom{- \sum\limits_{F = u,d,\ell ,\nu } \bigg[}+ {{\bar F}_f}\Big(\, {\dfrac{{m_f^F}}{v}{\delta _{fi}}s_{\beta\alpha} \!+\! \left( {\varepsilon _{fi}^F{P_R} + \varepsilon _{if}^{F*}{P_L}} \right)\!c_{\beta\alpha}} \Big)\!{F_i}{h^0}\\
	&\phantom{- \sum\limits_{F = u,d,\ell ,\nu } \bigg[} + i{\eta _F}{{\bar F}_f}\!\left( {\varepsilon _{fi}^F{P_R} - \varepsilon _{if}^{F*}{P_L}} \right)\!{F_i}{A^0} \bigg] \\
&  - \!\sqrt 2 \Big[ {{{\bar u}_f}\!\left( {V_{fj}^{}\varepsilon _{ji}^d{P_R} \! - \! \varepsilon _{jf}^{u*}V_{ji}^{{}}{P_L}} \right)\!{d_i}{H^ + } \!+\! {{\bar \nu }_f}\!\left( {U_{jf}^{{\rm{*}}}\varepsilon _{ji}^\ell {P_R}\! - \! \varepsilon _{jf}^{\nu *}U_{ij}^{{\rm{*}}}{P_L}} \right)\!{\ell _i}{H^ + } \! + \! {\rm h.c.}} \Big]. \label{Higgsbasis}
\end{split}
\end{align}
$V$ ($U$) is the CKM (PMNS) matrix, $m_i^F$ is the mass of the fermion $F=\{u,d,\ell,\nu\}$ with flavour index $i$ and 
\begin{equation}
-\eta_u=-\eta_\nu=\eta_\ell=\eta_d=1\,.
\end{equation}
We also allowed for the presence of right-handed neutrinos $N$ with a Majorana mass term $-1/2 M \bar N^c N$. This manifests itself in \eq{Higgsbasis} through the terms $m^\nu$ and $\varepsilon^\nu$ which otherwise would be absent. Note that $m^\nu$ corresponds to the Dirac mass term of the neutrinos which is related to the physical neutrino mass via the see-saw mechanism. Assuming a mass scale of the right-handed neutrinos at the TeV scale requires $m^\nu$ to be at most around 10 MeV. Thus we can safely neglect its effect on the Higgs couplings to fermions and focus on $\varepsilon^\nu$ which is decoupled from the neutrino masses and thus unconstrained.
\medskip  

We do not need to discuss the Higgs potential in detail since, in addition to the physical masses and mixing angles, only the two Higgs self-couplings enter in our calculation in the case of CP conservation. We will simply parametrize these couplings as $\lambda_{h_0 H^+ H^-}$ and $\lambda_{H_0 H^+ H^-}$ and refer the interested reader to Eq.~\eqref{HiggsselfcouplingDef} in the appendix for the explicit expressions.
\medskip

The Higgs basis defined in Eq. \eqref{Higgsbasis} is useful for calculations and phenomenology since fermion masses (generated from electroweak symmetry breaking) and the additional free couplings are decoupled. However, this basis is not motivated by a $\mathds{Z}_2$ symmetry which is capable to provide protection against flavour changing neutral currents. However, the parameters $\varepsilon^F_{ij}$ in the Higgs basis can be related to the ones within the four 2HDMs with natural flavour conservation (type-I/II/X/Y) as
\begin{equation}
\varepsilon_{ij}^{F}=c_{y}^{F}\dfrac{m_i^F}{v}\delta_{ij}+\frac{\tilde{\varepsilon}^{F}_{ij}}{c_{\tilde{\varepsilon}}^{F}}\,.
\end{equation}
The $\tilde{\varepsilon}^{F}_{ij}$ are the flavour changing entries in the new basis, i.e. the corrections to natural flavour conservation. The coefficients $c_{y}^{f}$ and $c_{\tilde{\varepsilon}}^{F}$ are given in Table~\ref{tablebasis}. In this basis, the terms $\tilde{\varepsilon}^{F}_{ij}$ break the $\mathds{Z}_2$ symmetry and lead to deviations from natural flavour conservation.

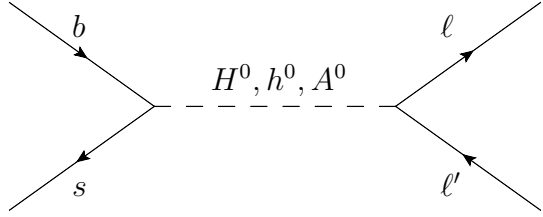
\begin{figure}[t]
	\begin{center}
		\begin{overpic}[scale=.60,,tics=10]
			{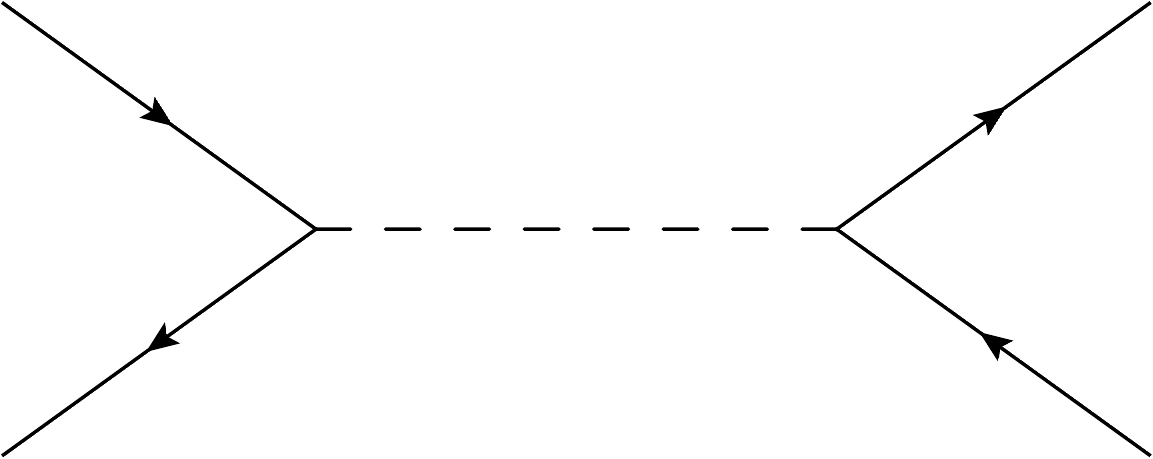}
			\put(12,33){$b$}
			\put(12,3){$s$}
			\put(38,23){$H^{0},h^{0},A^{0}$}
			\put(81,33){$\ell$}
			\put(81,3){$\ell^{\prime}$}
		\end{overpic}
	\end{center}
	\caption{Tree-level effects in $b\to s\ell^+\ell^-$ transitions induced by the flavour-changing couplings $\varepsilon^d_{23,32}$. These diagrams contribute to the Wilson coefficients of scalar operator $C_{S,P}^{(\prime)IJ}$ as given in Eq.~(\ref{bslltree}).}
	\label{bslltreeDiagram}
\end{figure}

\begin{boldmath}
\section{$b\to s\ell^+\ell^-$ Processes}
\label{bsll}
\end{boldmath}

We define the effective Hamiltonian giving direct effects in $b\to s \ell\ell^{(\prime)}$ and $b\to s\gamma$ transitions as
\begin{eqnarray}
H_{\rm eff}^{{\ell _I}{\ell _J}} &=&  - \dfrac{{4{G_F}}}{{\sqrt 2 }}{V_{tb}}V_{ts}^* \left(\sum\limits_{K=7,8} {C_K^{(\prime)} O_K^{(\prime)}} + \sum\limits_{K=9,10,S,P} {C_K^{(\prime)IJ}O_K^{(\prime)IJ}}\right)\,,
\end{eqnarray}
with the operators
\begin{align}
\begin{aligned}
O_7 &= \dfrac{{{e}}}{{16{\pi ^2}}} m_b \bar s{\sigma^{\mu\nu} }{P_R}bF_{\mu\nu}\,, &&& O_8 &= \dfrac{{{g_s}}}{{16{\pi ^2}}} m_b \bar s{\sigma^{\mu\nu} }T^a{P_R}b G_{\mu\nu}^a\,,\\
O_9^{IJ} &= \dfrac{{{e^2}}}{{16{\pi ^2}}}\bar s{\gamma_\mu }{P_L}b{{\bar \ell }_I}{\gamma ^\mu }{\ell _J}\,,&&& O_{10}^{IJ} &= \dfrac{{{e^2}}}{{16{\pi ^2}}}\bar s{\gamma _\mu }{P_L}b{{\bar \ell }_I}{\gamma ^\mu }{\gamma _5}{\ell _J}\,,\\
O_S^{IJ} &= \dfrac{{{e^2}}}{{16{\pi ^2}}}\bar s{P_L}b{{\bar \ell }_I}{\ell _J}\,, &&& O_P^{IJ} &= \dfrac{{{e^2}}}{{16{\pi ^2}}}\bar s{P_L}b{{\bar \ell }_I}{\gamma _5}{\ell _J}\,,
\end{aligned}
\end{align}
plus their primed counterparts which are obtained by exchanging $P_L$ and $P_R$. We did not include tensor operators here since they are not generated at the dim-6 level. 
\medskip

{In addition, we include four-quark operators which are generated by charged Higgs exchange (analogous to $O_2$ in the SM)}
\begin{align}
H_{\rm eff}^{sccb} &= - \dfrac{{4{G_F}}}{{\sqrt 2 }}{V_{tb}}V_{ts}^*\!\!\!\!\!\!\!\!\!\!\sum\limits_{K=\{LL,LR,RL,RR\}}^5 \!\!\!\!\!\!\!\!\!\!\!\!\! C_K O_K,
\label{sccboperators}
\end{align}
which can contribute to $b\to s\ell^+\ell^-$ processes at the loop-level. The operators are defined as
\begin{align}
\begin{aligned}
O_{AB} &= \left({{\bar s }}{P_A}c\right) \left(\bar c {P_B}b\right)\,,
\end{aligned}
\end{align}
with $A,B=L,R$ and the colour indices are contracted within the bilinears.
\medskip

\subsection{Tree-Level}

At tree-level, in the approximation of vanishing external momenta, we only get contributions to semi-leptonic scalar and pseudoscalar operators {from neutral Higgs exchange} (see Fig.~\ref{bslltreeDiagram}). They are given by
\begin{align}
\begin{split}
C_S^{IJ}=&\frac{16 \pi^2}{g_2^4 s_W^2 V_{tb}V_{ts}^{*}}\dfrac{m_W^2}{m_{H^\pm}^2}\varepsilon_{32}^{d*}\bigg(2s_{\beta\alpha}c_{\beta\alpha}\dfrac{m_{I}^{\ell}\delta_{IJ}}{v}(y_h-y_H)+L_{+}^{IJ}\bigg)\,,\\
C_P^{IJ} =& \frac{16\pi^2}{g_{2}^{4}s_W^2 V_{tb}V_{ts}^{*}}\dfrac{m_W^2}{m_{H^\pm}^2}\varepsilon_{32}^{d*}\bigg(\big(c_{\beta\alpha}^2 y_h+s_{\beta\alpha}^{2}y_H\big)\big(\varepsilon_{IJ}^{\ell}-\varepsilon_{JI}^{\ell*}\big)+y_A\big(\varepsilon_{IJ}^{\ell}+\varepsilon_{JI}^{\ell*}\big)\bigg)\,,\\
C_S^{\prime IJ} =&\frac{16\pi^2}{g_2^4 s_W^2 V_{tb}V_{ts}^{*}}\dfrac{m_W^2}{m_{H^\pm}^2}\varepsilon_{23}^{d}\bigg(2s_{\beta\alpha}c_{\beta\alpha}\dfrac{m_{I}^{\ell}\delta_{IJ}}{v}\big(y_{h}-y_{H}\big)-L_{-}^{IJ}\bigg) \,,\\
C_P^{\prime IJ} =&\frac{16\pi^2}{g_2^4 s_W^2 V_{tb}V_{ts}^{*}}\dfrac{m_W^2}{m_{H^\pm}^2}\varepsilon_{23}^{d}\bigg(\big(c_{\beta\alpha}^{2}y_{h}+s_{\beta\alpha}^2y_{H}\big)\big(\varepsilon_{IJ}^{\ell}-\varepsilon_{JI}^{\ell*}\big)-y_A\big(\varepsilon_{IJ}^{\ell}+\varepsilon_{JI}^{\ell*}\big)\bigg)\,,
\label{bslltree}
\end{split}
\end{align}
where we defined
\begin{align}
\begin{split}
L_{\pm}^{IJ}=y_A \big(
\varepsilon^\ell_{I J} - \varepsilon^{\ell *}_{J I}
\big)
\pm\left(c_{\beta\alpha}^2 y_h +s_{\beta\alpha}^2 y_H \right) \big(
\varepsilon^\ell_{I J} + \varepsilon^{\ell *}_{J I}
\big)\,,
\end{split}
\end{align}
and
\begin{align}
y_A=\dfrac{m_{H^{\pm}}^2}{m_{A_0}^2}\,,~~~~y_h=\dfrac{m_{H^{\pm}}^2}{m_{h_0}^2}\,,~~~~y_H=\dfrac{m_{H^{\pm}}^2}{m_{H_0}^2}\,.
\end{align}
In addition, we define for future convenience the squared mass ratios for heavy Majorana neutrino, up-type quark and the $W$ boson with respect to the charged Higgs 
\begin{align}
x_{i}=\frac{m_{N_i}^2}{m_{H^{\pm}}^2}\,,~~~~z_{i}=\frac{m_{u_i}^2}{m_{H^{\pm}}^2}\,,~~~~y=\frac{m_W^2}{m_{H^{\pm}}^2}\,.
\end{align}
\medskip

We derived \eq{bslltree} by working at leading order in the external momenta (which we will also do for all following results). This corresponds to an expansion in $m_{b,s}$ and $m_\ell$ over the Higgs masses which we assume to be at least at the EW scale. For consistency, one has to take into account all masses $m_{b,s}$ and $m_\ell$ in this expansion, also the ones entering via Higgs couplings\footnote{Note that it is a convenient feature of the Higgs basis that only the couplings which are related to EW symmetry breaking contain fermion masses (unlike in type-I/II/X/Y). Thus one can directly expand in these parameters without taking into account factors of $\sin\alpha$, $\tan\beta$, etc.}. Equation \eqref{bslltree} contains terms linear in light fermion masses which therefore correspond to dim-7 contributions. However, since from the expansion in the external momenta no dim-7 terms arise (the next non-vanishing order is dim-8), it is consistent to keep these terms even though in the loop effects, to be studied later, we only consider dim-6 terms. 
\medskip

The Wilson coefficients of the four-quark operators in \eq{sccboperators} due to tree-level charged Higgs exchange read
\begin{eqnarray}
\begin{aligned}
C_{LL}&=&\frac{4   \varepsilon_{k 2}^{d*} V^{*}_{2 k}  \varepsilon_{n 2}^{u*}V_{n 3}m_W^2}{g_2^2 V_{tb} V_{ts}^{*}m_{H^\pm}^2}\,,\\
C_{LR}&=&-\frac{4  V_{k 2}^{*} \varepsilon_{k 2}^{u} \varepsilon_{n 2}^{u*} V_{n 3}m_W^2}{g_2^2 V_{tb} V_{ts}^{*}m_{H^\pm}^2}\,,\\
C_{RL}&=&-\frac{4  \varepsilon_{k 2}^{d*} V_{2 k}^{*}  V_{2 n} \varepsilon_{n 3}^{d} m_W^2}{g_2^2 V_{tb} V_{ts}^{*}m_{H^\pm}^2}\,,\\
C_{RR}&=&\frac{4 V_{k 2}^{*} \varepsilon_{k 2}^{u} V_{2 n} \varepsilon_{n 3}^{d} m_W^2}{g_2^2 V_{tb} V_{ts}^{*}m_{H^\pm}^2}\,.
\label{qqqqtree}
\end{aligned}
\end{eqnarray}

\begin{boldmath}
\subsection{$b\to s\gamma$}
\end{boldmath}

\begin{figure}
	\begin{center}
	\begin{overpic}[scale=0.45,,tics=10]
		{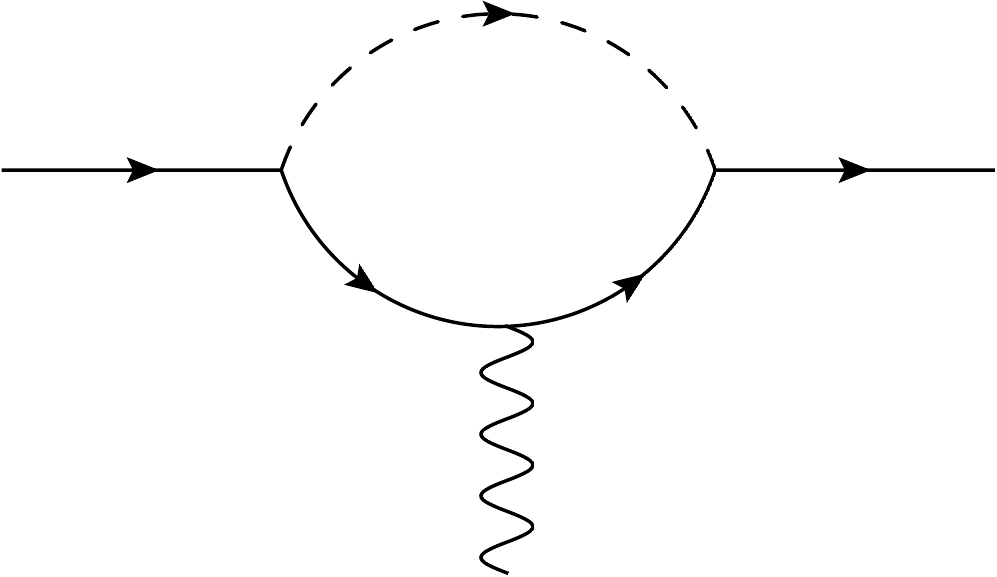}
		\put(10,45){$b$}
		\put(87,45){$s$}
		\put(46,47){$H^{-}$}
		\put(55,32){$c,t$}
		\put(35,32){$c,t$}
		\put(55,5){$\gamma$}
	\end{overpic}
	\begin{overpic}[scale=0.45,,tics=10]
		{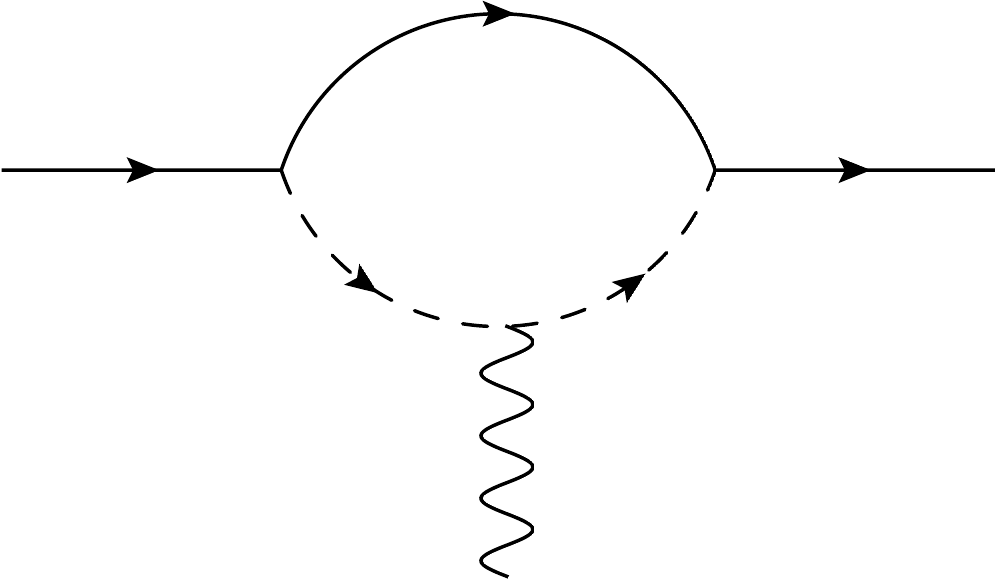}
		\put(10,45){$b$}
		\put(87,45){$s$}
		\put(43,47){$c,t$}
		\put(55,32){$H^{-}$}
		\put(36,32){$H^{-}$}
		\put(55,5){$\gamma$}
	\end{overpic}
	\begin{overpic}[scale=0.45,,tics=10]
		{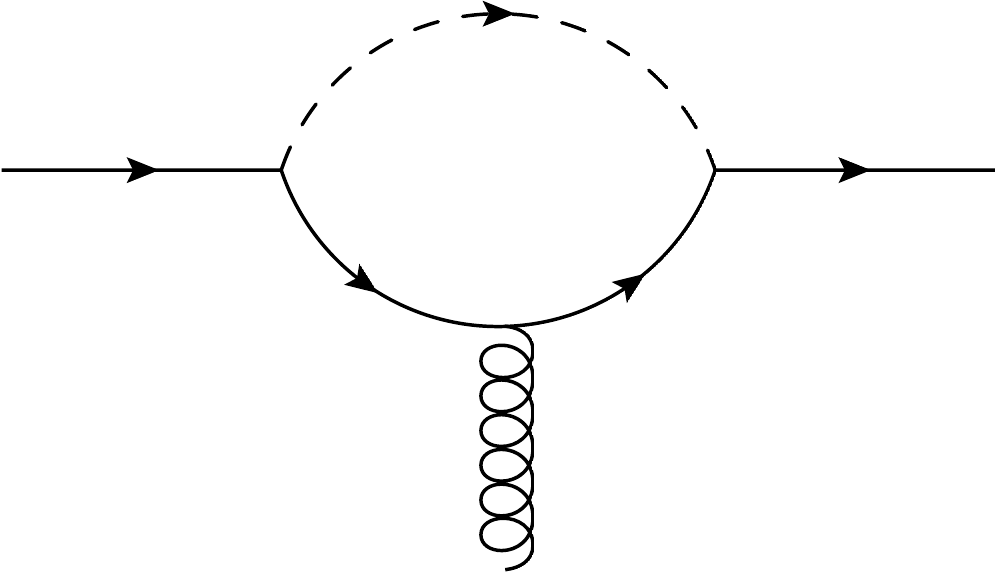}
		\put(10,45){$b$}
		\put(87,45){$s$}
		\put(46,47){$H^{-}$}
		\put(55,32){$c,t$}
		\put(35,32){$c,t$}
		\put(57,5){$g$}
	\end{overpic}
	\end{center}
	\caption{{Feynman diagrams showing the 2HDM contribution to $C_7^{(\prime)}$ and $C_8^{(\prime)}$ given in \eq{bsgamma1}, \eq{C8light} and \eq{C7soft}.}\label{bsgammadiagrams}}
\end{figure}

{Here (and for all loop effects to be calculated) we }do not consider multiple flavour changes which are phenomenologically known to be small. Regarding the (numerically) leading contributions due to the charged Higgs (see Fig.~\ref{bsgammadiagrams}) exchange we therefore only have to distinguish the top contribution (for which all particles in the loop are heavy) from the charm contribution (where we set the mass equal to zero). For the first case the result is given by
\begin{align}
\begin{split}
{C_7}^{H^\pm} =& - \dfrac{1}{{18}}\dfrac{m_W^2}{M_{H^\pm}^2}\dfrac{{V_{k2}^*\varepsilon _{k3}^u\varepsilon _{n3}^{u*}{V_{n3}}}}{{g_2^2{V_{tb}}V_{ts}^*}}f_{1}(z_3) -\dfrac{1}{3}\dfrac{{{m_{{t}}}}}{{{m_b}}}\dfrac{m_W^2}{M_{H^\pm}^2}\dfrac{{V_{k2}^*\varepsilon _{k3}^u{V_{3n}}\varepsilon _{n3}^d}}{{g_2^2{V_{tb}}V_{ts}^*}}f_{2}(z_3)\,,\\
{C^{\prime H^\pm}_7} =&- \dfrac{1}{{18}}\dfrac{m_W^2}{M_{H^\pm}^2}\dfrac{{\varepsilon _{k2}^{d*}V_{3k}^*{V_{3n}}\varepsilon _{n3}^d}}{{g_2^2{V_{tb}}V_{ts}^*}}f_{1}(z_3)  - \dfrac{1}{3}\dfrac{{{m_{{t}}}}}{{{m_b}}}\dfrac{m_W^2}{M_{H^\pm}^2}\dfrac{{\varepsilon _{k2}^{d*}V_{3k}^*\varepsilon _{n3}^{u*}{V_{n3}}}}{{g_2^2{V_{tb}}V_{ts}^*}}f_{2}(z_3)\,, \\
{C_8}^{H^\pm} =&  - \dfrac{1}{6}\dfrac{m_W^2}{M_{H^\pm}^2}\dfrac{{V_{k2}^*\varepsilon _{k3}^u\varepsilon _{n3}^{u*}{V_{n3}}}}{{g_2^2{V_{tb}}V_{ts}^*}}f_{3}(z_3) -\dfrac{{{m_{{t}}}}}{{{m_b}}}\dfrac{m_W^2}{M_{H^\pm}^2}\dfrac{{V_{k2}^*\varepsilon _{k3}^u{V_{3n}}\varepsilon _{n3}^d}}{{g_2^2{V_{tb}}V_{ts}^*}}f_{4}(z_3)\,,\\
{C^{\prime H^\pm}_8} =&  - \dfrac{1}{6}\dfrac{m_W^2}{M_{H^\pm}^2}\dfrac{{\varepsilon _{k2}^{d*}V_{3k}^*{V_{3n}}\varepsilon _{n3}^d}}{{g_2^2{V_{tb}}V_{ts}^*}}f_{3}(z_3)- \dfrac{{{m_{{t}}}}}{{{m_b}}}\dfrac{m_W^2}{M_{H^\pm}^2}\dfrac{{\varepsilon _{k2}^{d*}V_{3k}^*\varepsilon _{n3}^{u*}{V_{n3}}}}{{g_2^2{V_{tb}}V_{ts}^*}}f_{4}(z_3)\,,
\end{split}
\label{bsgamma1}
\end{align}
which is in agreement with e.g. \cite{Borzumati:1998tg,Borzumati:1998nx,Crivellin:2013wna}.
{Since we assume the charm quark in the denominator of the propagator to be massless, while we keep the leading term in the numerator, there is a dimensionally regularised infrared singularity which has to cancel with the EFT contribution originating from the four-quark operators defined in Eq.~(\ref{qqqqtree}). The result at the matching scale $\mu$ is thus given by}
\begin{align}
\begin{split}
{C_7}^{H^\pm}(\mu) =& - \dfrac{7}{{18}}\dfrac{m_W^2}{M_{H^\pm}^2}\dfrac{{V_{k2}^*\varepsilon _{k2}^u\varepsilon _{n2}^{u*}{V_{n3}}}}{{g_2^2{V_{tb}}V_{ts}^*}}- \dfrac{1}{3}\dfrac{{{m_{{c}}}}}{{{m_b}}}\dfrac{m_W^2}{M_{H^\pm}^2}\dfrac{{V_{k2}^*\varepsilon _{k2}^u{V_{2n}}\varepsilon _{n3}^d}}{{g_2^2{V_{tb}}V_{ts}^*}}\left(3+4\log\left(\dfrac{\mu^2}{m_{H^+}^2}\right)\right)\,,\\
{C^{\prime H^\pm}_7}(\mu) =& - \dfrac{7}{{18}}\dfrac{m_W^2}{M_{H^\pm}^2}\dfrac{{\varepsilon _{k2}^{d*}V_{2k}^*{V_{2n}}\varepsilon _{n3}^d}}{{g_2^2{V_{tb}}V_{ts}^*}} - \dfrac{1}{3}\dfrac{{{m_{{c}}}}}{{{m_b}}}\dfrac{m_W^2}{M_{H^\pm}^2}\dfrac{{\varepsilon _{k2}^{d*}V_{2k}^*\varepsilon _{n2}^{u*}{V_{n3}}}}{{g_2^2{V_{tb}}V_{ts}^*}}\left(3+4\log\left(\dfrac{\mu^2}{m_{H^+}^2}\right)\right)\,,\\
{C_8}^{H^\pm}(\mu) =&  - \dfrac{1}{3}\dfrac{m_W^2}{M_{H^\pm}^2}\dfrac{{V_{k2}^*\varepsilon _{k2}^u\varepsilon _{n2}^{u*}{V_{n3}}}}{{g_2^2{V_{tb}}V_{ts}^*}} -\dfrac{{{m_{{c}}}}}{{{m_b}}}\dfrac{m_W^2}{M_{H^\pm}^2}\dfrac{{V_{k2}^*\varepsilon _{k2}^u{V_{2n}}\varepsilon _{n3}^d}}{{g_2^2{V_{tb}}V_{ts}^*}}\left(3+2\log\left(\dfrac{\mu^2}{m_{H^+}^2}\right)\right)\,,\\
{C^{\prime H^\pm}_8}(\mu) =&  - \dfrac{1}{3}\dfrac{m_W^2}{M_{H^\pm}^2}\dfrac{{\varepsilon _{k2}^{d*}V_{2k}^*{V_{2n}}\varepsilon _{n3}^d}}{{g_2^2{V_{tb}}V_{ts}^*}}- \dfrac{{{m_{{c}}}}}{{{m_b}}}\dfrac{m_W^2}{M_{H^\pm}^2}\dfrac{{\varepsilon _{k2}^{d*}V_{2k}^*\varepsilon _{n2}^{u*}{V_{n3}}}}{{g_2^2{V_{tb}}V_{ts}^*}}\left(3+2\log\left(\dfrac{\mu^2}{m_{H^+}^2}\right)\right)\,.
\end{split}
\label{C8light}
\end{align}
The four fermion operators {in \eq{sccboperators} mix into $C_{7,8}^{(\prime)}$ (at order $\alpha_s^0$) from the matching $\mu$ down to the $B$ meson scale $\mu_b$, resulting in}
\begin{align}
\begin{split}
{C^{H^\pm}_{7\,\mathrm{mix}}}(\mu) =&  - \dfrac{4}{3}\dfrac{{{m_{{c}}}}}{{{m_b}}}\dfrac{m_W^2}{M_{H^\pm}^2}\dfrac{{V_{k2}^*\varepsilon _{k2}^u{V_{2n}}\varepsilon _{n3}^d}}{{g_2^2{V_{tb}}V_{ts}^*}}\log\left(\dfrac{\mu_{b}^2}{\mu^2}\right)
\,,\\
{C^{\prime H^\pm}_{7\,\mathrm{mix}}}(\mu) =&  - \dfrac{4}{3}\dfrac{{{m_{{c}}}}}{{{m_b}}}\dfrac{m_W^2}{M_{H^\pm}^2}\dfrac{{\varepsilon _{k2}^{d*}V_{2k}^*\varepsilon _{n2}^{u*}{V_{n3}}}}{{g_2^2{V_{tb}}V_{ts}^*}}\log\left(\dfrac{\mu_{b}^2}{\mu^2}\right)
\,,\\
{C^{H^\pm}_{8\,\mathrm{mix}}}(\mu) =&  \frac{3}{2} {C^{H^\pm}_{7\,\mathrm{mix}}}(\mu)\,,\\
{C^{\prime H^\pm}_{8\,\mathrm{mix}}}(\mu) =& \frac{3}{2} {C^{\prime H^\pm}_{7\,\mathrm{mix}}}(\mu)\,.
\end{split}
\label{C7soft}
\end{align}
Therefore, {the dependence on the matching scale $\mu$ cancels as required once both the hard matching contribution and the soft contribution from the EFT are added to each other. Since there is no constant term in \eq{C7soft} the inclusion of the soft contribution just leads to a replacement of $\mu$ by $\mu_b$ in Eq.~\eqref{C8light}. 
\medskip

While an explicit splitting into the hard matching contribution and the effect from the four-quark operators is necessary if one aims at including $\alpha_s$ corrections, this is not necessary at leading order and one can just add both contributions. In fact, since the neutral Higgs contribution is phenomenologically small, a leading order estimate is sufficient and we give here the sum of the soft and the hard contribution at the $B$ meson scale $\mu_b$}
\begin{equation}
\begin{aligned}
C^{H^0}_7\left(\mu_b\right)=&\frac{m_W^2 \varepsilon^d_{2 3}}{18 g_2^2 m_{H^+}^2 V^{*}_{t s} V_{t b}} \Big[
\varepsilon^{d *}_{3 3} \big(
y_A
+c_{\beta\alpha}^2 y_h
+s_{\beta\alpha}^2 y_H
\big)
+    3 \varepsilon^d_{3 3} \Big(
\left(3 + 2 \log\left(\frac{\mu_b^2}{m_{A_0}^2}\right) \right) y_A\\
&      - \left(3 + 2 \log\left(\frac{\mu_b^2}{m_{h_0}^2}\right)\right) c_{\beta\alpha}^2 y_h
-\left(3 + 2 \log\left(\frac{\mu_b^2}{m_{H_0}^2}\right)\right) s_{\beta\alpha}^2 y_H
\Big)
\Big]\,,\\
C_7^{\prime{H^0}}\left(\mu_b\right)=&\frac{m_W^2 \varepsilon^{d *}_{3 2} }{18 g_2^2 m_{H^+}^2 V^{*}_{t s} V_{t b}} \Big[
\varepsilon^d_{3 3} \big(
y_A
+c_{\beta\alpha}^2 y_h
+s_{\beta\alpha}^2 y_H
\big)
+3 \varepsilon^{d *}_{3 3} \Big(
\left(3 + 2 \log\left(\frac{\mu_b^2}{m_{A_0}^2}\right) \right) y_A\\
&      - \left(3 + 2 \log\left(\frac{\mu_b^2}{m_{h_0}^2}\right) \right) c_{\beta\alpha}^2 y_h
-\left(3 + 2 \log\left(\frac{\mu_b^2}{m_{H_0}^2}\right) \right) s_{\beta\alpha}^2 y_H
\Big)
\Big]\,,\\
C_8^{H^0}\left(\mu_b\right)=&-3C_7^{H_0}\left(\mu_b\right)\,,\\
C_8^{\prime H^0}\left(\mu_b\right)=&-3C_7^{\prime H_0}\left(\mu_b\right)\,.
\end{aligned}
\label{neutralHiggsC78}
\end{equation}

It is straightforward to use the NLO QCD corrections calculated in Ref.~\cite{Borzumati:1998tg} (for our prediction with a top-quark in the loop), where QCD corrections in a generic 2HDM with a discrete symmetry were considered. The Wilson coefficients $C_7$ and $C_8$ can be included by simply setting the couplings $X$ and $Y$ defined in Ref.~\cite{Borzumati:1998tg} to
\begin{align}
\begin{split}
\left|Y\right|^2 &= \frac{4 m_W^2}{g_2^2 m_t^2} \frac{V^*_{k2}\varepsilon^u_{k3}\varepsilon^{u*}_{l3}V_{l3}}{V_{33}V^*_{32}}\,,\\
X Y^* &= -\frac{4 m_W^2}{g_2^2 m_t m_b} \frac{V^*_{k2}\varepsilon^u_{k3}V_{3l}\varepsilon^{d}_{l3}}{V_{33}V^*_{32}}\,.
\end{split}
\end{align}
The primed operators can be treated in an analogous way taking into account that $C_2^{\prime}=0$.
\medskip

\begin{boldmath}
\subsection{One-Loop Effects in $b\to s\ell\ell^{(\prime)}$}
\end{boldmath}

We will now calculate the "leading" one-loop matching contributions to the operators $C_S^{(\prime)}$, $C_P^{(\prime)}$, $C_9^{(\prime)}$ and $C_{10}^{(\prime)}$. We will perform this calculation in a general $R_\xi$ gauge expanding all diagrams up to the first non-vanishing order in the external momenta, corresponding to dim-6 operators. In addition, we neglect all quark masses, except for the top-quark and integrate out all Higgses, $W$, $Z$ and the top at a common scale $m_{\rm EW}$. 
\medskip

By "leading" one-loop effects we also mean that we will only calculate the loop corrections to a Wilson coefficient if there is no corresponding tree-level effect. {In addition, we will neglect small effects originating from multiple flavour changes, i.e. $3\to 1\to 2$.} Thus, since the tree-level contribution involve $\varepsilon^d_{23,32}$, we will assume these couplings to be zero when calculating the loop correction. Therefore, flavour violation in the quark sector can either originate from the CKM matrix multiplying a diagonal $\varepsilon_{ii}^{d}$ or from the term $\varepsilon_{jf}^{u*}V_{ji}{P_L}$ which contributes both for diagonal and also off-diagonal elements $\varepsilon _{jf}^{u*}$. Note that the latter terms only enter via charged Higgs couplings to quarks. Hence, we just need to calculate diagrams with a charged Higgs and/or $W$ boson together with the corresponding charged Goldstones. Finally, we obtain gauge-invariant results. 
\medskip

\subsubsection{Self-Energies and Renormalization}

Here we will discuss the renormalization which can be solely derived from expressions for the self-energies. The reason is that in our setup (with $\varepsilon^d_{23,32}=0$) ultraviolet divergences only arise in (pseudo)scalar operators originating from Higgs penguins and Higgs couplings are intrinsically related to chirality changing self-energies (see Ref.~\cite{Crivellin:2010er}). We will also use this opportunity to illustrate the cancellation of the gauge dependence in the renormalization of the quark masses. We performed the calculation in a general $R_\xi$ gauge.
\medskip 

We begin by defining the self-energies as
\begin{equation}
\!\!\!\!\!\begin{gathered}\includegraphics{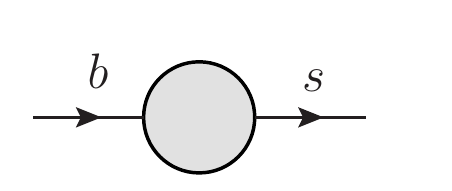}\end{gathered}\!\!\!\!\!\!\!\!\!\!\!= -i\left(\pslash P_L \Sigma^{LL}_{sb}+\pslash P_R \Sigma^{RR}_{sb}+P_R \Sigma^{LR}_{sb}+P_L \Sigma^{RL}_{sb}\right)\,,
\end{equation}
and we obtain the following expressions for $b\to s$ transitions
\begin{align}
\begin{split}
\Sigma^{LR}_{sb}=&\dfrac{e^2 V^{*}_{i 2} V_{i 3} m_b \xi z_i }{32 \pi^2 s_W^2 (z_i-\xi y) } \Big[\log(\xi y) -\log(z_i)\Big]\\
&-\dfrac{e^2 V^{*}_{i 2} V_{i 3} m_b z_i }{32 \pi^2 s_W^2 y } \Bigg[\log(z_i) -\left(1 + \dfrac{1}{\epsilon}+\log\left(\dfrac{\mu^2}{m_{H^+}^2}\right)\right)\Bigg]\\
&+\dfrac{\varepsilon^d_{3 3} V_{i 3} V^{*}_{k 2} \varepsilon^{u}_{k i} m_{u_i} }{8 \pi^2 }\Bigg[1 + \dfrac{1}{\epsilon}+\log\left(\dfrac{\mu^2}{m_{H^+}^2}\right)-\dfrac{\log\left(z_i\right) z_i}{z_i-1}\Bigg]\,,
\end{split}\\
\Sigma^{RL}_{sb}=&\dfrac{\varepsilon^{d *}_{2 2} \varepsilon^{u *}_{n i} V_{n 3} V^{*}_{i 2} m_{u_i} }{8 \pi^2 }\Bigg[1 + \dfrac{1}{\epsilon}+\log\left(\dfrac{\mu^2}{m_{H^+}^2}\right)-\dfrac{ \log\left(z_i\right) z_i}{z_i - 1}\Bigg]\,,
\end{align}
\begin{align}
\begin{split}
\Sigma^{LL}_{sb}=&-\dfrac{e^2 V^{*}_{i 2} V_{i 3}  z_i}{64 \pi^2 s_W^2 y}\left[\dfrac{1}{\epsilon}+\log\left(\dfrac{\mu^2}{m_{H^+}^2}\right)\right]-\dfrac{ V_{n 3} \varepsilon^{u *}_{n i} \varepsilon^{u}_{k i} V^{*}_{k 2}  }{16 \pi^2}\left[\dfrac{1}{\epsilon}+\log\left(\dfrac{\mu^2}{m_{H^+}^2}\right)\right]\\
&-\dfrac{e^2 V^{*}_{i 2} V_{i 3} \xi z_i}{16 \pi^2 s_W^2 (z_i-\xi y)}\Big[\log\left(\xi y\right) - \log\left(z_i\right)\Big] \\
&-\dfrac{e^2 V^{*}_{i 2} V_{i 3} z_i }{128 \pi^2 s_W^2 y\left(y - z_i\right)^{2}}\Big[6 \log\left(y\right) y^2+ 3( z_i^2-y^2) - \log\left(z_i\right)\left( 8  y^2 - 4 y z_i+ 2  z_i^2 \right)\Big]\\
&-\dfrac{ V_{n 3} \varepsilon^{u *}_{n i}\varepsilon^{u}_{k i} V^{*}_{k 2}  }{32 \pi^2\left(-1 + z_i\right)^{2}}\Big[1 - 4 z_i+ 3 z_i^2 - 2 \log\left(z_i\right) z_i^2 \Big]\,,
\end{split}\\
\Sigma^{RR}_{sb}=&\dfrac{\varepsilon^{d *}_{2 2} \varepsilon^d_{3 3} V^{*}_{i 2} V_{i 3}  }{16  \pi^2}z_i\Bigg[\dfrac{1}{1-z_i}+\dfrac{z_i\log(z_i)}{(z_i-1)^2}\Bigg],
\end{align}
with $\xi$ denoting the gauge parameter.
\medskip

Let us now consider the general effect of self-energies on kinetic terms and quark masses (see e.g. Ref.~\cite{Buras:2002vd}). First of all, one has to render the kinetic terms canonical, leading to the shifts in the quark fields
\begin{equation}
q_i^{L,R} \to \left( {\delta_{ij} + \dfrac{1}{2}\Sigma _{ij}^{LL,RR}} \right)q_j^{L,R}\,.\label{shift}
\end{equation}
These shifts then enter not only in all couplings but also in quark masses. Since the quark mass terms receive contributions from the chirality changing self-energies as well, we have
\begin{equation}
{m_f}{\delta _{fi}} \to m_{fi}^d = \left( {\delta_{fj} + \dfrac{1}{2}\Sigma _{fj}^{LL}} \right){{{m_j}}}{\delta _{jk}}\left( {\delta_{ki} + \dfrac{1}{2}\Sigma _{ki}^{RR}} \right) + \Sigma _{fi}^{LR}\,.
\end{equation}
The eigenvalues of this matrix {after renormalization in the $\overline{MS}$ scheme are identified with the physical quark masses, extracted from data according to the SM prescription.} Note that at first order in perturbation theory (i.e. linear in $\Sigma$), the eigenvalues just correspond to the diagonal terms
\begin{equation}
{m_i}\left( {1 + \dfrac{1}{2}\Sigma _{ii}^{RR} + \dfrac{1}{2}\Sigma _{ii}^{LL}} \right) + \Sigma _{ii}^{LR}\,,
\end{equation} 
where the dependence on $\xi$ drops out and thus rendering the renormalized parameter gauge-independent, as required for a physical quantity. The rotations that diagonalize the mass matrix as
\begin{equation}
U_{jf}^{L*}m_{jk}^dU_{ki}^R = m_i^d{\delta _{fi}}\,,
\end{equation}
read at leading order (considering only the $s$-$b$ sector)
\begin{align*}
U^L &\!=\!\!\left(\!\! {\begin{array}{*{20}{c}}
	1&{\dfrac{1}{2}\Sigma _{23}^{LL} + \dfrac{{\Sigma _{23}^{LR}}}{{{m_b}}}}\\
	{ - \dfrac{1}{2}\Sigma _{23}^{LL*} - \dfrac{{\Sigma _{23}^{LR*}}}{{{m_b}}}}&1
	\end{array}} \!\!\right)\!, ~~
U^R \!=\!\! \left(\!\! {\begin{array}{*{20}{c}}
	1&{\dfrac{1}{2}\Sigma _{23}^{RR} + \dfrac{{\Sigma _{23}^{RL}}}{{{m_b}}}}\\
	{ - \dfrac{1}{2}\Sigma _{23}^{RR*} - \dfrac{{\Sigma _{23}^{RL*}}}{{{m_b}}}}&1
	\end{array}} \!\!\right)\!.
\end{align*}
These rotations, together with the shifts in \eq{shift} result in
\begin{equation}
\tilde U_{}^L \approx \left( {\begin{array}{*{20}{c}}
{1 + \dfrac{1}{2}\Sigma _{22}^{LL}}&{\Sigma _{23}^{LL} + \dfrac{{\Sigma _{23}^{LR}}}{{{m_b}}}}\\
{ - \dfrac{{\Sigma _{23}^{LR*}}}{{{m_b}}}}&{1 + \dfrac{1}{2}\Sigma _{33}^{LL}}
\end{array}} \right),\;\;\;\tilde U_{}^R \approx \left( {\begin{array}{*{20}{c}}
{1 + \dfrac{1}{2}\Sigma _{22}^{RR}}&{\Sigma _{23}^{RR} + \dfrac{{\Sigma _{23}^{RL}}}{{{m_b}}}}\\
{ - \dfrac{{\Sigma _{23}^{RL*}}}{{{m_b}}}}&{1 + \dfrac{1}{2}\Sigma _{33}^{RR}}
\end{array}} \right)\,.\label{Utilde}
\end{equation}
This agrees with the diagrammatical approach of Ref.~\cite{Crivellin:2010gw} and confirms the statements of Ref.~\cite{Logan:2000iv} that diagrams involving flavour changing self-energies can be treated as one-particle irreducible. Thus, we apply \eq{Utilde} to the couplings $\varepsilon^d_{ij}$ and take into account all self-energy contributions.
\medskip

Let us now turn to the renormalization. As stated above, it can be determined solely from the expressions for the self-energies. Unlike in the SM or in 2HDMs with natural flavour conservation, our results for $b\to s\ell^+\ell^-$ will be divergent for generic couplings $\varepsilon^u_{ij}$. The reason for this is that once $\varepsilon^u_{ij}$ does not correspond to a special case of the four 2HDMs with natural flavour conservation (see Table~\ref{tablebasis}), the $\mathds{Z}^2$ symmetry in the Yukawa sector is broken and no symmetry protects $\varepsilon^d_{ij}$ from being flavour changing. In fact, counterterms to off-diagonal elements of $\varepsilon^d_{ij}$ are required to render the result finite. Since all divergences originate from Higgs penguin diagrams, we can determine the $1/\epsilon$ structure of our results from the self-energies. For this, we start with the interaction basis in which the Yukawa Lagrangian is given by
\begin{equation}
 - L_Y^{EW} = {\bar d_f}\left( {Y_{fi}^dH_0^d + \tilde{\varepsilon} _{fi}^dH_0^u} \right){P_R}{d_i} + {\bar u_f}\left( {Y_{fi}^uH_0^u + \tilde{\varepsilon} _{fi}^uH_0^d} \right){P_R}{u_i}\,,
 \label{eq:interaction_basis}
\end{equation}
where for simplicity we considered the neutral current part only. Assuming \eqref{eq:interaction_basis} is already in the basis with diagonal mass matrices, the masses then are given by
\begin{align}
m_{fj}^d \delta_{ji} &= {v_d}Y_{fi}^d + {v_u}\tilde{\varepsilon}_{fi}^d\,, & m_{fi}^u &= {v_u}Y_{fi}^u + {v_d}\tilde{\varepsilon}_{fi}^u\,.
\end{align}
Since the chirality flip on the fermion line in $\Sigma _{23}^{LR}$ always originates from an up-quark mass, we can define 
\begin{align}
\left( {Y_{kl}^{u*}{v_u} + \tilde{\varepsilon}_{kl}^{u*}{v_d}} \right)\sigma _{fi}^{kl} = {\left. {\Sigma _{fi}^{LR}} \right|_{\rm div}}\,.
\end{align}
We keep only the relevant divergent part and we obtain
\begin{align}
\sigma _{23}^{ij}& =   \dfrac{{\tilde{\varepsilon}_{33}^d V_{k2}^*\tilde{\varepsilon}_{ki}^u{V_{j3}}}}{{8{\pi ^2}}}\dfrac{1}{\epsilon}\,, & \sigma _{32}^{ij} &=  -\dfrac{{\tilde{\varepsilon} _{22}^dV_{k3}^*\tilde{\varepsilon} _{ki}^u{V_{j2}}}}{{8{\pi ^2}}}\dfrac{1}{\epsilon}\,.
\end{align}
We invert the relations in Table~\ref{tablebasis} to go to the Higgs basis and set for consistency reasons the quark masses to zero. Then we apply the rotations in \eq{Utilde} and find
\begin{align}
\begin{aligned}
\delta \varepsilon _{23}^d &= {\left( {\frac{{\Sigma _{23}^{LR}}}{{{m_b}}}\varepsilon _{33}^d - \varepsilon _{22}^d\left( {\Sigma _{23}^{RR} + \frac{{\Sigma _{23}^{RL}}}{{{m_b}}}} \right)} \right)_{\rm div}} - \sigma _{sb}^{ij}\varepsilon _{ji}^{u*}\,,\\
\delta \varepsilon _{32}^d &= {\left( {\varepsilon _{33}^d\frac{{\Sigma _{23}^{RL*}}}{{{m_b}}} - \left( {\Sigma _{23}^{LL*} + \frac{{\Sigma _{23}^{LR*}}}{{{m_b}}}} \right)\varepsilon _{22}^d} \right)_{\rm div}} - \sigma _{bs}^{ij}\varepsilon _{ji}^{u*}\,,
\end{aligned}
\label{counterterms}
\end{align}
where the definition for the bare couplings $\varepsilon_{23,32}^{d(0)}=\varepsilon_{23,32}^{d}+\delta\varepsilon_{23,32}^{d}$ was used. Again, note that these counterterms are independent of the gauge parameter $\xi$. As we will see later, these counterterms, inserted into the tree-level expressions for $B_s\to\ell^+\ell^-$ (see \eq{bslltree}), will render the results finite.
\medskip

\subsubsection{$Z$ and $\gamma$ Penguins}

\begin{figure}
	\begin{center}
	\begin{overpic}[scale=0.5,,tics=10]
		{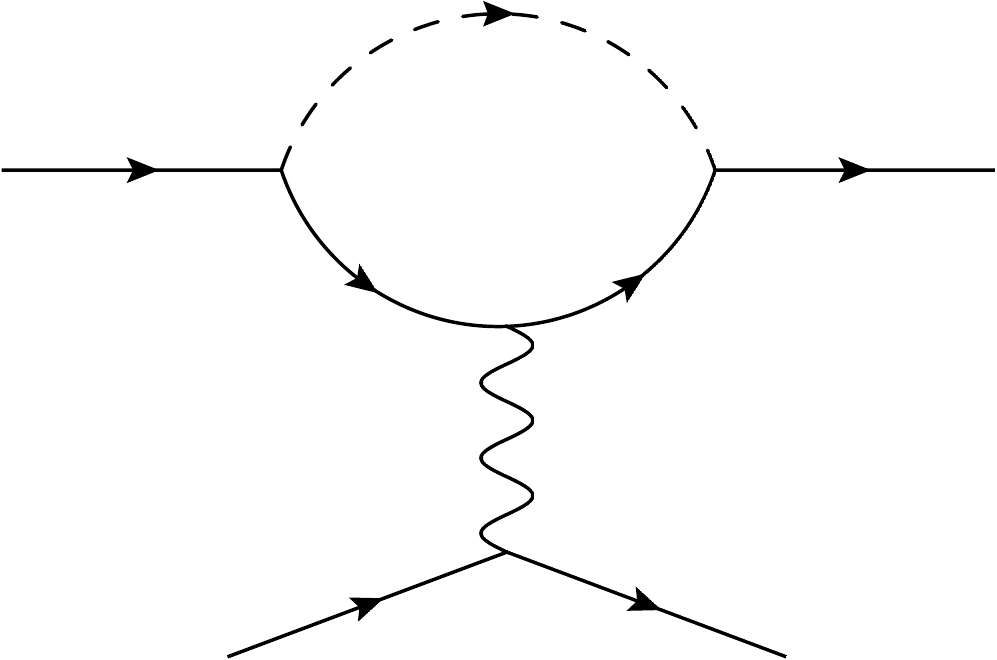}
		\put(10,53){$b$}
		\put(87,53){$s$}
		\put(46,56){$H^{-}$}
		\put(55,40){$c,t$}
		\put(35,40){$c,t$}
		\put(55,20){$Z,\gamma$}
		\put(25,7){$\ell^{\prime}$}
		\put(75,7){$\ell$}
	\end{overpic}
	\begin{overpic}[scale=0.5,,tics=10]
		{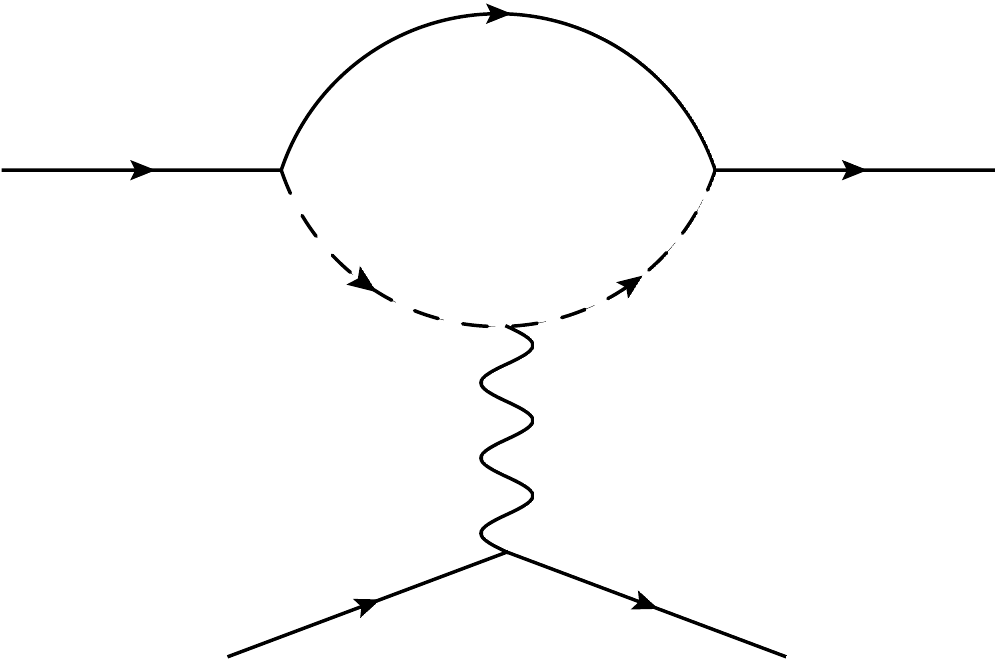}
		\put(10,53){$b$}
		\put(87,53){$s$}
		\put(45,56){$c,t$}
		\put(56,40){$H^{-}$}
		\put(36,40){$H^{-}$}
		\put(55,20){$Z,\gamma$}
		\put(25,7){$\ell^{\prime}$}
		\put(75,7){$\ell$}
	\end{overpic}
	\end{center}
	\caption{Feynman diagrams showing the off-shell photon and $Z$ penguin contributions to $C^{(\prime)}_{9(10)}$, given in Eqs.~(\ref{gammaZ}, \ref{C9top}, \ref{light_quark_gamma_pen})}
	\label{ZgammaPengDiagram}
\end{figure}
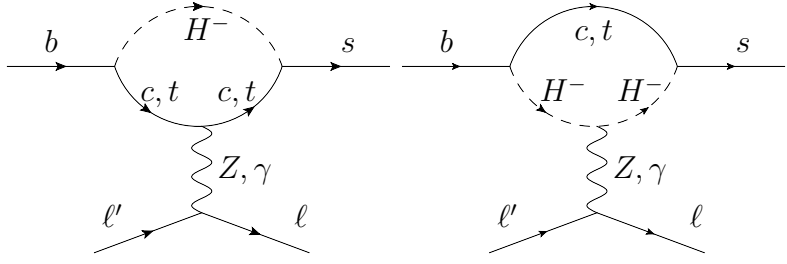

The Wilson coefficients originating from $Z$ penguins and involving the charged Higgs (see Fig.~\ref{ZgammaPengDiagram}), are only relevant for top exchange and are given by
\begin{align}
\begin{split}
C_9^{IJ} &=-\delta_{IJ}\dfrac{V_{k2}^*\varepsilon _{k3}^u{\varepsilon _{n3}^{u*}{V_{n3}}}}{2e^2{{V_{tb}}V_{ts}^*}}\left( {1 - 4s_W^2} \right)\big(I_{1}(z_3)-1\big)\,,\\
C_{10}^{IJ} &=\delta_{IJ}\dfrac{{V_{k2}^*\varepsilon _{k3}^u}\varepsilon _{n3}^{u*}{V_{n3}}}{2{e^2}{{V_{tb}}V_{ts}^*}}\big(I_{1}(z_3)-1\big)\,,\\
C_9^{\prime IJ} &= \delta_{IJ}\dfrac{{\varepsilon _{k2}^{d*}V_{3k}^*{V_{3n}\varepsilon _{n3}^d}\,}}{2e^2{{V_{tb}}V_{ts}^*}}\left( {1 - 4s_W^2} \right)\big(I_{1}(z_3)-1\big)\,,\\
C_{10}^{\prime IJ} &= -\delta_{IJ}\dfrac{\varepsilon _{k2}^{d*}V_{3k}^*{{V_{3n}}\varepsilon _{n3}^d\,}}{{2{e^2}{V_{tb}}V_{ts}^*}}\big(I_{1}(z_3)-1\big)\,,
\label{gammaZ}
\end{split}
\end{align}
where the loop function $I_{1}(x)$ is defined in the appendix. Note that $I_{1}(0)-1=0$ justifying that we only consider the top quark here.
\medskip

For the off-shell photon penguin, also shown in Fig.~\ref{ZgammaPengDiagram}, we obtain for the top quark
\begin{align}
\begin{split}
C_9^{IJ} &= \delta_{IJ}\dfrac{{V_{k2}^*\varepsilon _{k3}^u\varepsilon _{n3}^{u*}{V_{n3}}}}{27{g_2^2{V_{tb}}V_{ts}^*}} \dfrac{m_W^2}{M_{H^\pm}^2} f_5(z_{3})\,,\\
C_9^{\prime IJ} &=\delta_{IJ} \dfrac{{\varepsilon _{k2}^{d*}V_{3k}^*{V_{3n}}\varepsilon _{n3}^d}}{27{g_2^2{V_{tb}}V_{ts}^*}} \dfrac{m_W^2}{M_{H^\pm}^2}f_{5}(z_3)\,.
\end{split}
\label{C9top}
\end{align}
Concerning light-quarks, the hard matching contributions get amended by the mixing of the four-quark operators in \eq{qqqqtree} into $C_9$ and $C_9^\prime$. We obtain
\begin{align}
\begin{split}
C_9^{IJ}(\mu_b) &=\delta_{IJ}\dfrac{2}{{27}}\dfrac{{V_{k2}^*\varepsilon _{k2}^u\varepsilon _{n2}^{u*}{V_{n3}}}}{{g_2^2{V_{tb}}V_{ts}^*}} \dfrac{m_W^2}{M_{H^\pm}^2}\Bigg(19+12 \log\! \left(\!\dfrac{\mu_{b}^2}{M_{H^\pm}^2}\!\right) \Bigg)\,,\\
C_9^{\prime IJ}(\mu_b) &=\delta_{IJ} \dfrac{2}{{27}}\dfrac{{\varepsilon _{k2}^{d*}V_{2k}^*{V_{2n}}\varepsilon _{n3}^d}}{{g_2^2{V_{tb}}V_{ts}^*}}\dfrac{m_W^2}{M_{H^\pm}^2}\Bigg(19+12 \log\! \left(\!\dfrac{\mu_{b}^2}{M_{H^\pm}^2}\!\right) \Bigg)\,.
\end{split}
\label{light_quark_gamma_pen}
\end{align}
The same result can be obtained by expanding \eq{C9top} in $m_t$ and then replacing $m_t$ in the logarithm by the $B$ meson scale $\mu_{b}$. Once more, note that at LO adding the soft to the hard matching contribution is justified.
\medskip

\subsubsection{Higgs Penguin and $W$-Higgs Boxes}

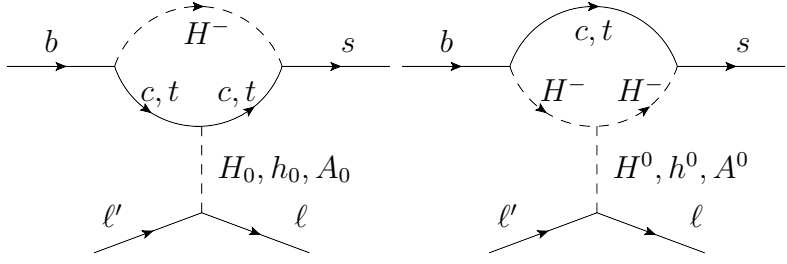
\begin{figure}
	\begin{center}
	\begin{overpic}[scale=0.5,,tics=10]
		{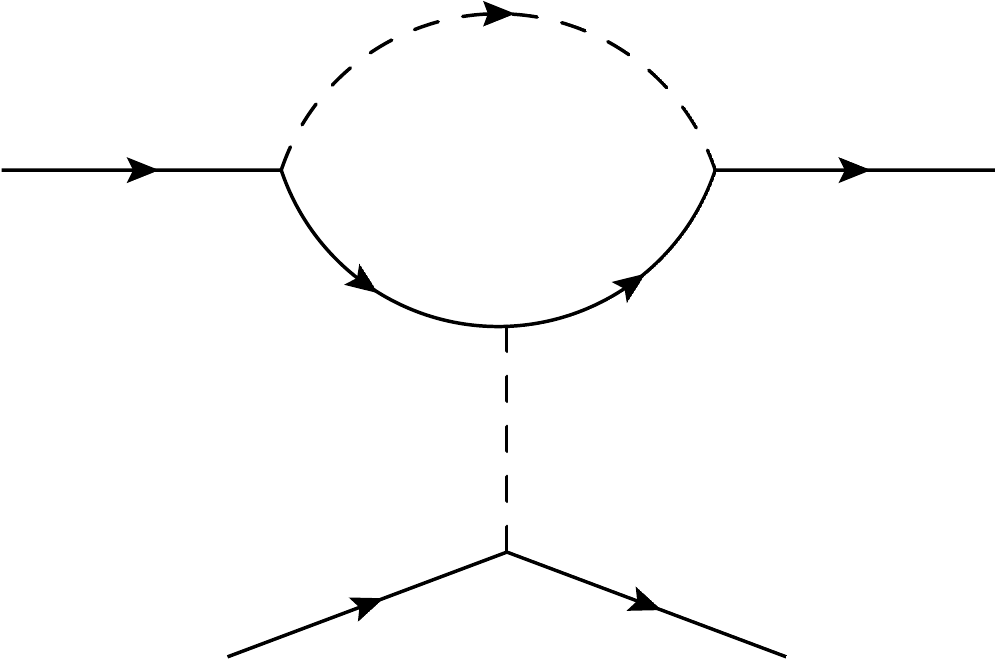}
		\put(10,53){$b$}
		\put(87,53){$s$}
		\put(46,55){$H^{-}$}
		\put(55,40){$c,t$}
		\put(35,40){$c,t$}
		\put(55,20){$H_{0},h_{0},A_{0}$}
		\put(25,7){$\ell^{\prime}$}
		\put(75,7){$\ell$}
	\end{overpic}
	\begin{overpic}[scale=0.5,,tics=10]
		{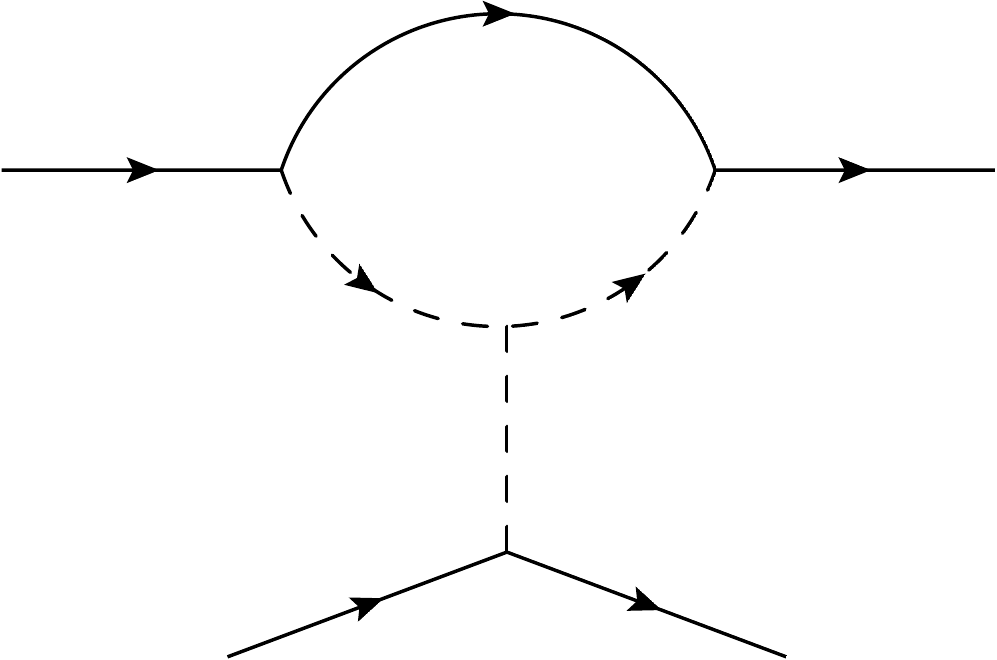}
		\put(10,53){$b$}
		\put(87,53){$s$}
		\put(45,56){$c,t$}
		\put(56,40){$H^{-}$}
		\put(36,40){$H^{-}$}
		\put(55,20){$H^{0},h^{0},A^{0}$}
		\put(25,7){$\ell^{\prime}$}
		\put(75,7){$\ell$}
	\end{overpic}

	\end{center}
	\caption{Higgs-penguin Feynman diagrams contributing to $C_{S(P)(HH)}^{(\prime) IJ}$ in Eqs.~(\ref{HiggsPenguRes}, \ref{HiggsPenguRes2}).}
	\label{HiggsPengDiagramm}
\end{figure}

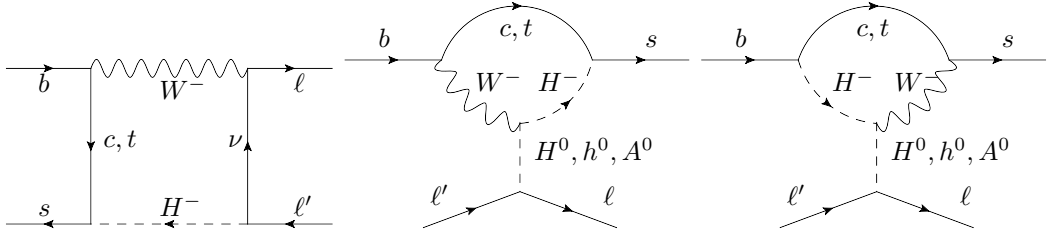
\begin{figure}[t]
	\begin{center}
		\begin{overpic}[scale=.45,,tics=10]
			{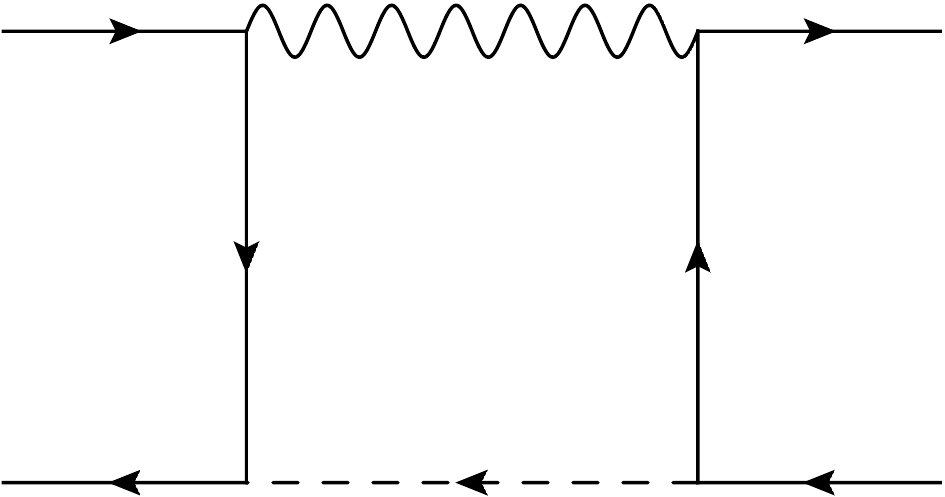}
			\put(10,42){\footnotesize$b$}
			\put(10,4){\footnotesize$s$}
			\put(30,25){\footnotesize$c,t$}
			\put(68,25){\footnotesize$\nu$}
			\put(47,40){\footnotesize$W^{-}$}
			\put(47,4){\footnotesize$H^{-}$}
			\put(88,4){\footnotesize$\ell^{\prime}$}
			\put(88,42){\footnotesize$\ell$}
		\end{overpic}
		\begin{overpic}[scale=0.45,,tics=10]
			{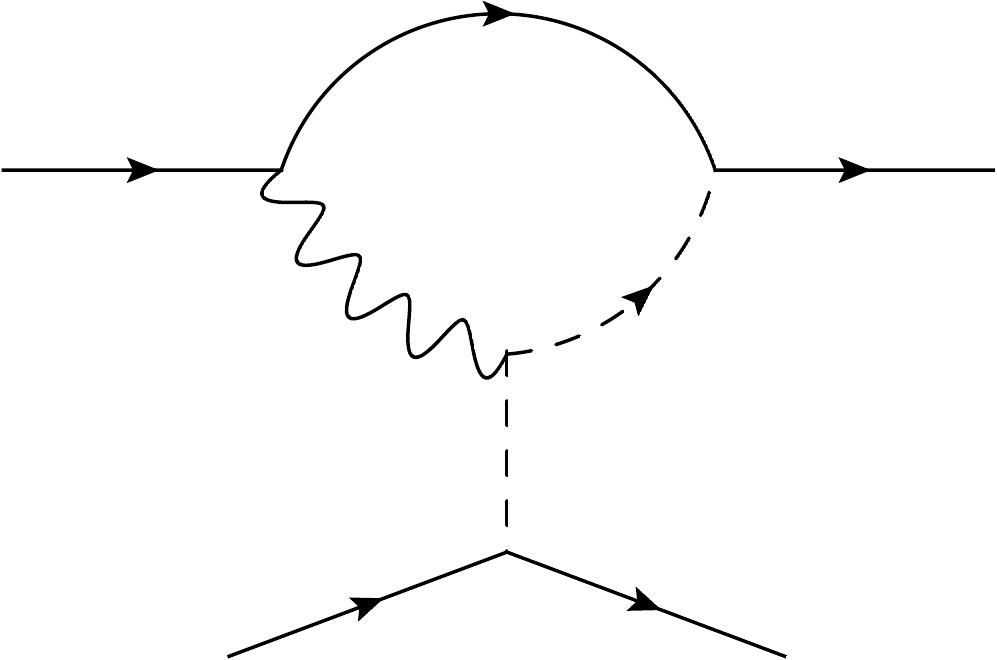}
			\put(10,53){\footnotesize$b$}
			\put(87,53){\footnotesize$s$}
			\put(45,56){\footnotesize$c,t$}
			\put(56,40){\footnotesize$H^{-}$}
			\put(38,40){\footnotesize$W^{-}$}
			\put(55,20){\footnotesize$H^{0},h^{0},A^{0}$}
			\put(25,7){\footnotesize$\ell^{\prime}$}
			\put(75,7){\footnotesize$\ell$}
		\end{overpic}
		\begin{overpic}[scale=0.45,,tics=10]
			{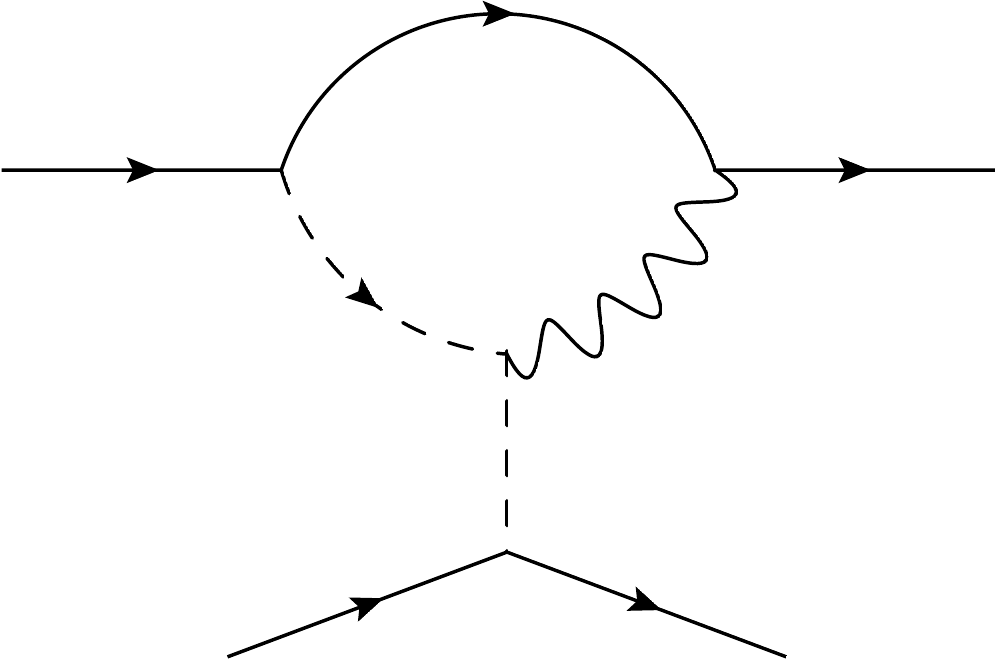}
			\put(10,53){\footnotesize$b$}
			\put(87,53){\footnotesize$s$}
			\put(45,56){\footnotesize$c,t$}
			\put(56,40){\footnotesize$W^{-}$}
			\put(38,40){\footnotesize$H^{-}$}
			\put(55,20){\footnotesize$H^{0},h^{0},A^{0}$}
			\put(25,7){\footnotesize$\ell^{\prime}$}
			\put(75,7){\footnotesize$\ell$}
		\end{overpic}
	\end{center}
	\caption{Mixed $H$-$W$ box-diagrams and Higgs penguins contributing to $C_{S(P)(HW)}^{(\prime)IJ}$ in Eq.~\ref{CSPHW}). It is understood for the $W$ diagrams that the Goldstone bosons are implicitly included. }
	\label{HWDiagrams}
\end{figure}

Here, contributions originating from flavour changing self-energies appear that are parametrically enhanced by 
\begin{equation}
t_i=\dfrac{m_{u_i}}{m_b}\,,
\end{equation}
for $i=3$. Using these definitions, the neutral Higgs penguin contributions involving a top quarks and a $H^\pm$ in the loop, (see Fig.~\ref{HiggsPengDiagramm}) read
\begin{align}
\begin{split}
C_{S(HH)}^{IJ}=&\dfrac{\varepsilon^{d *}_{2 2}}{g_2^4 s_W^2 V^{*}_{t s} V_{t b}}\Bigg(\!\!\!-\dfrac{m_W^2}{2 m_{H^\pm}^2} L_{+}^{IJ} \Big[
     4 I_1\left(z_3\right) t_3 ( z_3-1) \big(
       \varepsilon^d_{3 3} V^{*}_{k 2} \varepsilon^{u}_{k 3} V_{3 3}
      -\varepsilon^{d *}_{3 3} V^{*}_{3 2} \varepsilon^{u *}_{n 3} V_{n 3}
  \big)\\
&     -2 \log\left(\dfrac{\mu^2}{m_{H^+}^2}\right) \big(
      2 \left(\varepsilon^d_{3 3} V^{*}_{k 2} \varepsilon^{u}_{k 3} V_{3 3}-\varepsilon^{d *}_{3 3} V^{*}_{3 2} \varepsilon^{u *}_{n 3} V_{n 3}\right) t_3
       + 2 V^{*}_{3 2} \varepsilon^u_{3 3} \varepsilon^{u *}_{n 3} V_{n 3}\\
&       - V^{*}_{k 2} \varepsilon^u_{k 3} \varepsilon^{u *}_{n 3} V_{n 3}
  \big)
    -I_0\left(z_3\right) V^{*}_{k 2} \varepsilon^{u}_{k 3} \varepsilon^{u *}_{n 3} V_{n 3}
      +4 I_5\left(z_3, z_3\right) V^{*}_{3 2} \varepsilon^u_{3 3} \varepsilon^{u *}_{n 3} V_{n 3}
\Big]\\
&+
  2 I_4\left(z_3, z_3\right) V^{*}_{3 2} \varepsilon^{u *}_{3 3} \varepsilon^{u *}_{n 3} V_{n 3} L_{-}^{IJ} \dfrac{m_W^2}{m_{H^\pm}^2}
\\
&-{V^{*}_{3 2} \varepsilon^{u *}_{n 3} V_{n 3} {\dfrac{m_W}{m_{H^\pm}}} \sqrt{z_3} \big(
       \varepsilon^\ell_{I J}
      +\varepsilon^{\ell *}_{J I}
  \big)} \Big[
     2 (1 - I_1\left(z_3\right)) c_{\beta\alpha} g_2 s_{\beta\alpha} (y_h-y_H)\\
&   +I_1\left(z_3\right) {\dfrac{m_W}{m_{H^\pm}}} \Big( c_{\beta\alpha} y_h \dfrac{\lambda_{h_0 H^+ H^-}}{m_{H^+}}  
       - s_{\beta\alpha} y_H \dfrac{\lambda_{H_0 H^+ H^-}}{m_{H^+}} \Big)
\Big]\Bigg)\,,\\
C_{S(HH)}^{\prime IJ}\!=&\dfrac{1}{g_2^4 s_W^2 V^{*}_{t s} V_{t b}}\!\Bigg(\! \dfrac{m_W^2}{m_{H^\pm}^2} L_{-}^{IJ}\!\Big[
     \!-\!2 I_1\!\left(z_3\right) t_3 ( z_3\!-\!1) \big(\!
       \left(\varepsilon^d_{3 3}\right)^{\!2}  V^{*}_{k 2} \varepsilon^{u}_{k 3} V_{3 3}
      \!-\!\varepsilon^{d *}_{2 2} \varepsilon^d_{2 2} V^{*}_{3 2} \varepsilon^{u *}_{n 3} V_{n 3}
  \big)\\
&    +\!2 \log\left(\dfrac{\mu^2}{m_{H^+}^2}\!\right)\!\big(\!\!
       -\!\varepsilon^d_{3 3} V^{*}_{k 2} \varepsilon^{u}_{k 3} \varepsilon^{u *}_{3 3} V_{3 3}
      \!+\!\left(\!\left(\varepsilon^d_{3 3}\right)^{\!2}\! V^{*}_{k 2} \varepsilon^{u}_{k 3} V_{3 3}\!-\!\varepsilon^{d *}_{2 2} \varepsilon^d_{2 2} V^{*}_{3 2} \varepsilon^{u *}_{n 3} V_{n 3}\!\right)\! t_3
  \big)\\
&      +\varepsilon^d_{3 3} \big(
       I_7\left(z_3\right) \varepsilon^{d *}_{2 2} \varepsilon^d_{2 2} V^{*}_{3 2} V_{3 3}
      +2 I_5\left(z_3, z_3\right) V^{*}_{k 2} \varepsilon^{u}_{k 3} \varepsilon^{u *}_{3 3} V_{3 3}
  \big)
\Big]\\
&
  -2 I_4\left(z_3, z_3\right) \varepsilon^d_{3 3} V^{*}_{k 2} \varepsilon^{u}_{k 3} \varepsilon^u_{3 3} V_{3 3} L_{+}^{IJ} \dfrac{m_W^2}{m_{H^\pm}^2}
\\
&-{\varepsilon^d_{3 3} V^{*}_{k 2} \varepsilon^{u}_{k 3} V_{3 3} {\dfrac{m_W}{m_{H^\pm}}} \sqrt{z_3} \big(
       \varepsilon^\ell_{I J}
      +\varepsilon^{\ell *}_{J I}
  \big)} \Big[
     2 (1 - I_1\left(z_3\right)) c_{\beta\alpha} g_2 s_{\beta\alpha} (y_h - y_H)\\
&  +I_1\left(z_3\right) {\dfrac{m_W}{m_{H^\pm}}} \Big( c_{\beta\alpha} \dfrac{\lambda_{h_0 H^+ H^-}}{m_{H^+}} y_h
      - \dfrac{\lambda_{H_0 H^+ H^-}}{m_{H^+}} s_{\beta\alpha}  y_H \Big)
\Big]\Bigg)\,.
\end{split}
\label{HiggsPenguRes}
\end{align}
The charm contribution is obtained in the limit $z\to0$ and is explicitly given in the appendix. The top quark contributions of diagrams including both $W^\pm$ and $H^\pm$, i.e. mixed boxes and Higgs penguins with a $W$ in the loop (see Fig.~\ref{HWDiagrams}) yield the result
\begin{align}
\begin{aligned}
C_{S(HW)}^{IJ}=&\dfrac{\varepsilon^{d *}_{2 2} }{g_2^2 s_W^2 }\left(
    \dfrac{z_3}{4}\log\left(\dfrac{\mu^2}{m_{H^+}^2}\right) L_{+}^{IJ}
  + \dfrac{1}{8}I_3\left(y, z_3\right) L_{+}^{IJ}
  + I_2\left( z_3\right) \varepsilon^\ell_{I J}
\right)\,,\\
C_{S(HW)}^{\prime IJ}=&\dfrac{\varepsilon^d_{3 3} }{g_2^2 s_W^2 }\left(
    \dfrac{z_3}{2}\log\left(\dfrac{\mu^2}{m_{H^+}^2}\right)  L_{-}^{IJ}
  - \dfrac{1}{2}I_6\left(z_3\right) L_{-}^{IJ}
  + I_2\left( z_3\right) \varepsilon^{\ell *}_{J I}
\right)\,,
\label{CSPHW}
\end{aligned}
\end{align}
which constitutes a gauge invariant subset. The expressions for $C_P^{(\prime) IJ}$ are related to the ones given above by 
\begin{align}
C_P^{IJ}&=C_S^{IJ}\Big|_{\varepsilon^{\ell *}_{J I}\to-\varepsilon^{\ell *}_{J I}}\,,&
C_P^{\prime IJ}&=C_S^{\prime I J}\Big|_{\varepsilon^{\ell *}_{J I}\to-\varepsilon^{\ell *}_{J I}}.
\end{align}
The charm contribution vanishes in limit $m_c\to 0$ since the loop functions involved approach zero in the approximation.
\medskip

The sum of the results in \eq{HiggsPenguRes} and \eq{CSPHW} is renormalized in the $\overline{MS}$ scheme using the counterterms of \eq{counterterms} inserted into the tree-level expressions of \eq{bslltree}. As a further check of the correctness of the result, note that in the limit of one of the four 2HDMs with natural flavour violation the result is finite without any counterterm.
\medskip

\subsubsection{$H^\pm$ Boxes}

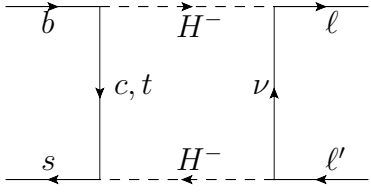
\begin{figure}
	\begin{center}
		\begin{overpic}[scale=.50,,tics=10]
			{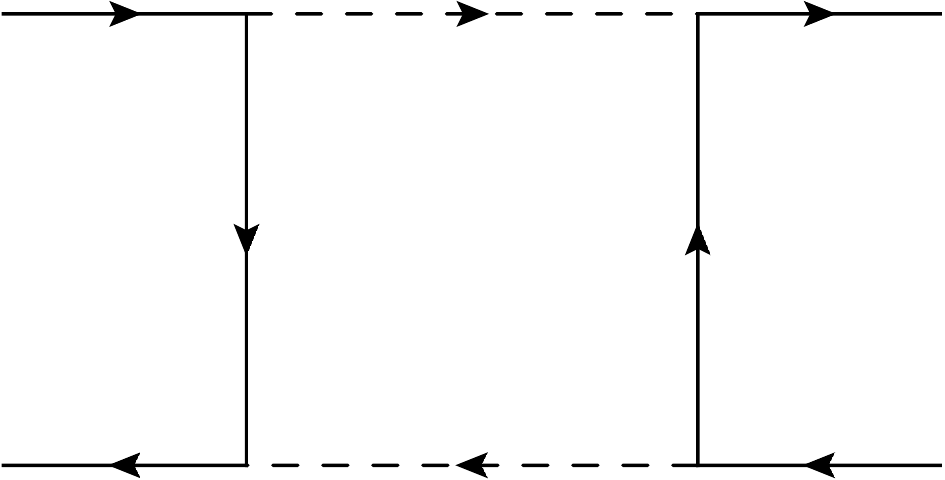}
			\put(10,42){$b$}
			\put(10,4){$s$}
			\put(30,25){$c,t$}
			\put(68,25){$\nu$}
			\put(47,41){$H^{-}$}
			\put(47,4){$H^{-}$}
			\put(88,4){$\ell^{\prime}$}
			\put(88,42){$\ell$}
		\end{overpic}
	\end{center}
	\caption{Box diagrams involving only charged Higgses contributing to $C_{9,10}^{(\prime)IJ}$ in Eq.~(\ref{HHboxes}).}
	\label{HHDiagram}
\end{figure}

The expressions for the box diagrams involving two charged Higgses (see Fig.~\ref{HHDiagram}) are given by
\begin{eqnarray}
\begin{aligned}
C_9^{IJ}  &= \frac{-m_W^2}{g_2^4 s_W^2 V_{tb}V_{ts}^{*}m_{H^\pm}^2}\big( {V_{k2}^*\varepsilon _{ki}^u\varepsilon _{ni}^{u*}{V_{n3}}} \big)\big( \varepsilon_{mI}^{\ell *}\varepsilon_{mJ}^{\ell}I_{1}(z_{i})-{{U_{Ip}}\varepsilon _{pj}^\nu \varepsilon _{mj}^{\nu *}U_{Jm}^*} I_{8}(z_i,x_j)\big)\,,\\
C_{10}^{IJ}  &= \frac{-m_W^2}{g_2^4 s_W^2 V_{tb}V_{ts}^{*}m_{H^\pm}^2}\big( {V_{k2}^*\varepsilon _{ki}^u\varepsilon _{ni}^{u*}{V_{n3}}} \big)\big(\varepsilon_{mI}^{\ell *}\varepsilon_{mJ}^{\ell}I_{1}(z_i)+ {{U_{Ip}}\varepsilon _{pj}^\nu \varepsilon _{mj}^{\nu *}U_{Jm}^*} I_{8}(z_i,x_j\big)\big)\,,\\
C_9^{\prime IJ}  &=\frac{-m_W^2}{g_2^4 s_W^2 V_{tb}V_{ts}^{*}m_{H^\pm}^2}\big( {\varepsilon _{k2}^{d*}V_{ik}^*{V_{in}}\varepsilon _{n3}^d} \big)\big( {\varepsilon _{mI}^{\ell *}\varepsilon _{mJ}^\ell }I_{1}(z_i)-{{U_{Ip}}\varepsilon _{pj}^\nu \varepsilon _{mj}^{\nu *}U_{Jm}^*}I_{8}(z_i,x_j\big)\big)\,,\\
C_{10}^{\prime IJ}  &= \frac{-m_W^2}{g_2^4 s_W^2 V_{tb}V_{ts}^{*}m_{H^\pm}^2}\big( {\varepsilon _{k2}^{d*}V_{ik}^*{V_{in}}\varepsilon _{n3}^d}\big)\big( {\varepsilon _{mI}^{\ell *}\varepsilon _{mJ}^\ell }I_{1}(z_i)+{{U_{Ip}}\varepsilon _{pj}^\nu \varepsilon _{mj}^{\nu *}U_{Jm}^*} I_{8}(z_i,x_j\big)\big)\,.
\label{HHboxes}
\end{aligned}
\end{eqnarray}
Note that $\varepsilon^\ell$ ($\varepsilon^\nu$) generates $C_{9}=(-)C_{10}$ and $C_{9}^{\prime}=(-)C_{10}^{\prime}$. The limit $m_c\to0$ exists and the corresponding expressions for the loop-functions are given in the appendix.
\medskip

\subsection{Processes and Observables}

For $b\to s\mu^+\mu^-$ transitions it is helpful to distinguish three regimes, the one of scalar operators ($C_S^{(\prime)}$ and $C_P^{(\prime)}$), the one of vector operators ($C_9^{(\prime)}$ and $C_{10}^{(\prime)}$) and the one of magnetic operators ($C_{7}^{(\prime)}$). In $B_s\to\ell\ell^{\prime}$ processes both scalar and vector operators enter in the branching ratio (see e.g.~\cite{Dedes:2008iw,Crivellin:2013wna})
\begin{align}
\begin{split}
{{\rm Br}}&\left[ B_s \to \ell_{I}^{+} \ell_{J}^{-}
\right] = \dfrac{G_{F}^{4} M_{W}^{4} s_W^4}{32 \pi^{5}} \big| V_{tb}^* V_{ts}\big|^2
f\left(r_{I}^{2},r_{J}^{2}\right) \,  M_{B_s} \,  f_{B_s}^{2} \,
\left(m_{\ell_{I}}+m_{\ell_{J}}\right)^{2}   \,   \tau_{B_s}   \\
& \times \left\{       \left|  \dfrac{M_{B_s}^{2}
	\left(C_{P}^{IJ*}-C_{P}^{\prime IJ*}\right)}{\left(m_{q_{f}}+m_{q_{i}}\right)\left(m_{\ell_{I}}+m_{\ell_{J}}\right)} 
-\left(C_{10}^{IJ*}-C_{10}^{\prime IJ*}\right)  \right|^{2} \!\!
\left[1-(r_{I}-r_{J})^{2}\right]  \right.   \\
&+ \left. \left|  \dfrac{M_{B_s}^{2} \left(C_{S}^{\prime IJ*} -C_{S}^{IJ*}\right)}{(m_{q_{f}}+m_{q_{i}})(m_{\ell_{I}}+m_{\ell_{J}})} 
+\dfrac{(m_{\ell_{I}}-m_{\ell_{J}})}{(m_{\ell_{I}}+m_{\ell_{J}})}\left(C_9^{IJ*}-C_9^{\prime IJ *}\right)
\right| ^{2} \!\!\left[1-(r_{I}+r_{J})^{2}\right]    \right\} \,,
\label{BRmostgen}
\end{split}
\end{align}
with $f\left(r_I,r_J\right)$ and $r_I$ defined as
\begin{equation}
f\left(r_I,r_J\right)=\sqrt{1-2\left(r_I+r_J\right)+\left(r_I-r_J\right)^2}\,\,,\quad r_I=\frac{m_{\ell_I}}{M_{B_s}}.
\end{equation}
Note that \cite{Crivellin:2013wna} uses a different definition for the operator basis. As one can see, the effect of scalar operators is enhanced by a factor $\approx M_{B_s}^2/(m_b m_{\ell_{{\rm max}[I,J]}})$, with respect to the vector ones. Thus, these processes (also since they are two-body decays) are most sensitive to scalar operators taking into account \eq{BsmumuEXP} and \eq{BsmumuSM}. However, the effect of vector operators cannot be neglected here, since they have different parametric dependences, notably contributions independent of $\varepsilon^d_{ij}$.
\medskip

Concerning magnetic operators, the inclusive $b\to s\gamma$ decay is most sensitive. The SM prediction~\cite{Misiak:2015xwa,Misiak:2017woa} 
\begin{equation}
{\rm Br}[B\to X_s\gamma]_{\rm SM}=(3.36 \pm 0.23) \times {10}^{ -4 }
\end{equation}
has to be compared to the experimental value~\cite{Amhis:2016xyh}
\begin{equation}
{\rm Br}[B\to X_s\gamma]_{\rm EXP}=(3.32 \pm 0.15) \times 10^{ -4 }\,.
\end{equation}
In case of vanishing $C_{7,8}^\prime$ one can use the numerical formula~\cite{Misiak:2015xwa} to express the branching ratio in terms of the Wilson coefficients\footnote{For a more detailed analysis included primed operators see e.g. Ref.~\cite{Hurth:2003dk}.} at the matching scale
\begin{equation}
{\rm Br}[B\to X_s\gamma]=(3.36 \pm 0.23-8.22 C_7-1.99 C_8) \times 10^{-4}\,.
\end{equation}
{Note that the contributions in Eqs.~(\ref{C8light}, \ref{neutralHiggsC78}), which would require the addition of the four Fermion operators in Eq.~(\ref{qqqqtree}) are all proportional to $\varepsilon^d$, which we set to zero in our analysis.}
Finally, semi-leptonic decays are important to constrain vector operators since their dependence on scalar ones is very weak~\cite{Becirevic:2012fy}. However, many processes and observables have been measured and one therefore should use a global fit to constrain $C_{9,10}^{(\prime)\mu\mu}$ (taking also into account $B_s\to \mu^+\mu^-$ if one assumes the absence of scalar operators). The scenario with a lepton flavour conserving $C_{10}$ effect ($C_{10}^U$) and a contribution to $C_9=-C_{10}$ with muons only ($C_9^V=-C_{10}^V$) (following the conventions of Ref.~\cite{Alguero:2018nvb}) is phenomenologically the most important scenario for us. We will discuss this in the next section.
\medskip
 
Concerning the case of decays into tau leptons, one can calculate the semi-leptonic processes using the relevant expressions for the factors. We use the results of Ref.~\cite{Capdevila:2017iqn} and find for tau leptons
\begin{align}\label{eq:NPdep1}
\begin{split}
10^7\times{\rm{Br}}&\left[{{B}\to K\tau^+\tau^-}\right]^{[15,22]}= \Big(1.20+0.15\,C_9^\prime-0.42\,C_{10}^\prime+0.02\,C_9^{\prime\,2}\\ &+0.05\,C_{10}^{\prime\,2}+0.15\,C_9^\text{NP}-0.42\,C_{10}^\text{NP}+0.04\,C_9^\text{NP}C_9^\prime+0.10\,C_{10}^\text{NP}C_{10}^\prime\\
&+0.02\,C_9^{\text{NP}\,2}+0.05\,C_{10}^{\text{NP}\,2}\Big) \pm
\Big(0.12+0.02\,C_9^\text{NP}-0.04\,C_{10}^\text{NP}\\
&+0.01\,C_9^\prime-0.04\,C_{10}^\prime+0.08\,C_{10}^{\prime\,2} +0.01\,C_{10}^\text{NP}C_{10}^\prime+0.01\,C_{10}^{\text{NP}\,2}\Big)\,,\\
\end{split}
\end{align}
\begin{align}
\begin{split}
10^7\times{\rm{Br}}&\left[{{B}\to K^*\tau^+\tau^-}\right]^{[15,19]} =\Big(0.98-0.30\,C_9^\prime+0.12\,C_{10}^\prime+0.05\,C_9^{\prime\,2}\\
&+0.02\,C_{10}^{\prime\,2}+0.38\,C_9^\text{NP}-0.14\,C_{10}^\text{NP}-0.08\,C_9^\text{NP}C_9^\prime-0.03\,C_{10}^\text{NP}C_{10}^\prime\\ &+0.05\,C_9^{\text{NP}\,2}+0.02\,C_{10}^{\text{NP}\,2}\Big) \pm \Big(0.09+0.03\,C_9^\text{NP}-0.01\,C_{10}^\text{NP}\\
&-0.01\,C_9^\text{NP}C_9^\prime -0.03\,C_9^\prime-0.01\,C_9^\prime C_{10}^\prime +0.01\,C_9^{\prime\,2}-0.01\,C_{10}^{\prime\,2}\Big)\,,
\end{split}
\end{align}
\begin{align}
\begin{split}
10^7\times&{\rm{Br}}\left[{B_s}\to \phi\tau^+\tau^{-}\right]^{[15,18.8]}=\Big(0.86-0.28\,C_9^\prime+0.10\,C_{10}^\prime+0.05\,C_9^{\prime\,2}\\
&+0.01\,C_{10}^{\prime\,2}+0.34\,C_9^\text{NP}-0.11\,C_{10}^\text{NP}-0.08\,C_9^\text{NP}C_9^\prime-0.02\,C_{10}^\text{NP}C_{10}^\prime\\
&+0.05\,C_9^{\text{NP}\,2}+0.01\,C_{10}^{\text{NP}\,2}\Big) \pm \big(0.06+0.02\,C_9^\text{NP}-0.02\,C_9^\prime+0.02\,C_{10}^{\prime\,2}\big)\,.
\end{split}
\label{eq:NPdep3}
\end{align}
For lepton flavour violating transitions one finds~\cite{Crivellin:2015era}
\begin{align}
\Br[B\to K\ell^+\ell^{\prime-}] &= 10^{-9} \left(a_{K\ell\ell^\prime}
\left|C_9^{\ell\ell^\prime} + C_9^{\prime\ell\ell^\prime} \right|^2 +
b_{K\ell\ell^\prime}\left|C_{10}^{\ell\ell^\prime} +
C_{10}^{\prime\ell\ell^\prime} \right|^2\right)\,,
\label{bktaumu}
\end{align}
\begin{align}
\begin{split}
\Br[B\to K^*\ell^+\ell^{\prime-}] &= 10^{-9} \left(a_{K^*\ell\ell^\prime}
\left|C_9^{\ell\ell^\prime} + C_9^{\prime\ell\ell^\prime} \right|^2 +
b_{K^*\ell\ell^\prime}\left|C_{10}^{\ell\ell^\prime} +
C_{10}^{\prime\ell\ell^\prime} \right|^2 \right.  \\ 
&~~+\left.
c_{K^*\ell\ell^\prime}\left|C_9^{\ell\ell^\prime}
-C_9^{\prime\ell\ell^\prime} \right|^2 +
d_{K^*\ell\ell^\prime}\left|C_{10}^{\ell\ell^\prime}
-C_{10}^{\prime\ell\ell^\prime} \right|^2 \right)\,,
\label{bkstaumu}
\end{split}
\end{align}

with
\begin{center}
	\begin{tabular}{|c|c|c|c|c|c|c|}
		\hline
		$\ell\ell^\prime $ & $a_{K\ell\ell^\prime}$ & $b_{K\ell\ell^\prime}$ &
		$a_{K^*\ell\ell^\prime}$ & $b_{K^*\ell\ell^\prime}$ &
		$c_{K^*\ell\ell^\prime}$ & $d_{K^*\ell\ell^\prime}$ \\
		\hline
		$\;\tau\mu\;$ & $\;9.6 \pm 1.0\;$ & $\;10.0 \pm 1.3\;$ & $\;3.0 \pm
		0.8\;$ & $\;2.7 \pm 0.7\;$ & $\;16.4 \pm 2.1\;$ & $\;15.4 \pm 1.9\;$
		\\
		$\mu e$ & $15.4 \pm 3.1$ & $15.7 \pm 3.1$ & $5.6 \pm 1.9$ & $5.6 \pm
		1.9$ & $29.1 \pm 4.9$ & $29.1 \pm 4.9$ \\
		\hline
	\end{tabular}
\end{center}
\medskip

\begin{boldmath}
\section{$b\to s\nu\bar{\nu}$, $B_s-\bar B_s$ Mixing, $a_\mu$ and $\ell\to\ell^\prime\gamma$}
\label{calculation2}
\end{boldmath}
Let us now turn to the matching for the remaining $b\to s$ processes, $b\to s\nu\bar{\nu}$ and $B_s-\bar B_s$ mixing. In addition, we consider the anomalous magnetic moments of charged leptons together with the closely related radiative lepton decays and $h\to\tau\mu$.
\medskip

\begin{boldmath}
\subsection{$b\to s\nu\bar{\nu}$}
\end{boldmath}
For $b\to s\nu\bar{\nu}$ processes the corresponding effective Hamiltonian is defined as
\begin{eqnarray}
H_{eff}^{{\nu _I}{\nu _J}} &=&  - \dfrac{{4{G_F}}}{{\sqrt 2 }}{V_{tb}}V_{ts}^*\left( {C_L^{IJ}O_L^{IJ} + C_R^{IJ}O_R^{IJ}} \right)\,,
\end{eqnarray}
with the operators
\begin{align}
O_L^{IJ} &=\dfrac{{{e^2}}}{{16{\pi ^2}}}\bar s{\gamma _\mu }{P_L}b{{\bar \nu }_I}{\gamma ^\mu }\left( {1 - {\gamma _5}} \right){\nu _J}\,,
&&& O_R^{IJ} &= \dfrac{{{e^2}}}{{16{\pi ^2}}}\bar s{\gamma _\mu }{P_R}b{{\bar \nu }_I}{\gamma ^\mu }\left( {1 - {\gamma _5}} \right){\nu _J}\,.
\end{align}
From box diagrams with charged Higgses we obtain
\begin{align}
C_L^{IJ} 
&=\frac{y}{g_{2}^4 s_W^2 V_{tb}V_{ts}^{*}}\big(V_{m2}^*\varepsilon _{mi}^u\varepsilon _{li}^{u*}{V_{l3}}U_{nI}^*\varepsilon _{nj}^\ell \varepsilon _{pj}^{\ell *}{U_{pJ}}\big)I_{1}(z_i)\,,\\
C_R^{IJ} 
&=\frac{y}{g_{2}^4 s_W^2 V_{tb}V_{ts}^{*}}\big(\varepsilon _{m2}^{d*}V_{im}^*{V_{il}}\varepsilon _{l3}^dU_{nI}^*\varepsilon _{nj}^\ell \varepsilon _{pj}^{\ell *}{U_{pJ}}\big)I_{1}(z_i)\,.
\end{align}
We follow \cite{Buras:2014fpa} and define 
\begin{equation}
\epsilon_{IJ}=\frac{\sqrt{\big|C_L^{IJ}\big|^2+\big|C_R^{IJ}\big|^2}}{\big|C_L^{SM}\big|},\qquad \eta_{IJ}=\frac{-{\rm Re} \left[C_L^{IJ}C_R^{JI*}\right]}{\big|C_L^{IJ}\big|^2+\big|C_R^{IJ}\big|^2}\,.
\end{equation}
This allows us to write the branching ratio in terms of
\begin{equation}
R_K=\frac{1}{3}\sum_{\{I,J\}=1}^{3}(1-2\eta_{IJ})\epsilon_{IJ}^2,\qquad R_{K^*}=\frac{1}{3}\sum_{\{I,J\}=1}^{3}(1+\kappa_\eta \eta_{IJ})\epsilon_{IJ}^2\,,
\end{equation}
where $\kappa_\eta$ encapsules the dependence on the form factors. In Ref.~\cite{Buras:2014fpa} this quantity is evaluated using as input for the $B \to K^*$ form factors a combined fit to lattice and LCSR results performed in \cite{Straub:2015ica}, finding $\kappa_\eta=1.34\pm0.04$. The branching ratio reads
\begin{equation}
\mathrm{Br}\left[B\to X_s \nu \bar \nu\right]\approx \mathrm{Br}\left[B\to X_s \nu \bar \nu\right]_{\rm{SM}}\left(\frac{\kappa_\eta R_K+2R_K^*}{2+\kappa_\eta}\right)\,.
\end{equation}
This has to be compared to the experimental limits~\cite{Grygier:2017tzo} 
\begin{equation}
{R_K^{\nu\bar{\nu}}} < 3.9\,,\qquad  {R_{{K^*}}^{\nu\bar{\nu}}} < 2.7\,.
\end{equation}
\medskip

\subsection{$B_s-\bar B_s$ Mixing}
\begin{figure}[t]
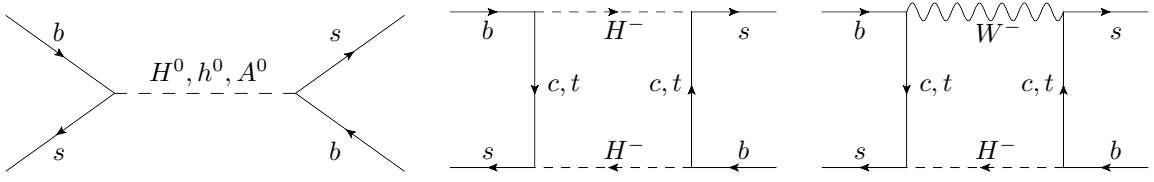

	\centering
	\begin{overpic}[scale=.45,,tics=10]
		{tree-level.pdf}
		\put(12,33){\footnotesize$b$}
		\put(12,3){\footnotesize$s$}
		\put(36,23){\footnotesize$H^{0},h^{0},A^{0}$}
		\put(81,33){\footnotesize$s$}
		\put(81,3){\footnotesize$b$}
	\end{overpic}
	\hspace{0.3cm}
	\begin{overpic}[scale=.45,,tics=10]
		{box_HH.pdf}
		\put(10,41){\footnotesize$b$}
		\put(10,4){\footnotesize$s$}
		\put(30,25){\footnotesize$c,t$}
		\put(61,25){\footnotesize$c,t$}
		\put(47,41){\footnotesize$H^{-}$}
		\put(47,4){\footnotesize$H^{-}$}
		\put(88,4){\footnotesize$b$}
		\put(88,41){\footnotesize$s$}
	\end{overpic}
	\hspace{0.3cm}
	\begin{overpic}[scale=.45,,tics=10]
		{box_HW.pdf}
		\put(10,41){\footnotesize$b$}
		\put(10,4){\footnotesize$s$}
		\put(30,25){\footnotesize$c,t$}
		\put(61,25){\footnotesize$c,t$}
		\put(47,40){\footnotesize$W^{-}$}
		\put(47,4){\footnotesize$H^{-}$}
		\put(88,4){\footnotesize$b$}
		\put(88,41){\footnotesize$s$}
	\end{overpic}
\caption{Feynman diagrams contributing to $B_{s}-\bar{B}_{s}$ mixing. Note that the tree-level contribution is absent for $\varepsilon_{23}^{d}=\varepsilon_{32}^{d}=0$.}
\label{fig:Bs-mixing}
\end{figure}

The effective Hamiltonian is defined as
\begin{align}
H_{\rm eff}^{\Delta F=2}=\sum_{a=1}^{5}C_{a}O_{a}+\sum_{a=1}^{3}C_{a}^{\prime}O_{a}^{\prime}\,,
\end{align}
with
\begin{align}
\begin{aligned}
O_{1}^{(\prime)}&=\left[\bar{s}_{\alpha}\gamma^{\mu}P_{L(R)}b_{\alpha}\right]\left[\bar{s}_{\beta}\gamma_{\mu}P_{L(R)}b_{\beta}\right]\,, &&&
O_{2}^{(\prime)}&=\left[\bar{s}_{\alpha} P_{L(R)}b_{\alpha}\right]\left[\bar{s}_{\beta} P_{L(R)}b_{\beta}\right]\,,\\
O_{3}^{(\prime)}&=\left[\bar{s}_{\alpha} P_{L(R)}b_{\beta}\right]\left[\bar{s}_{\beta} P_{L(R)}b_{\alpha}\right]\,,&&&
O_{4}&=\left[\bar{s}_{\alpha} P_{L}b_{\alpha}\right]\left[\bar{s}_{\beta} P_{R}b_{\beta}\right]\,,\\
O_{5}&=\left[\bar{s}_{\alpha} P_{L}b_{\beta}\right]\left[\bar{s}_{\beta} P_{R}b_{\alpha}\right]\,.
\end{aligned}
\end{align}
We obtain at tree level (see left diagram in Fig. \ref{fig:Bs-mixing})
\begin{align}
\begin{split}
C_{2}&=-\frac{1}{2}\big(\varepsilon_{32}^{d*}\big)^2\bigg(\frac{s_{\beta\alpha}^{2}}{m_{H_0}^2}+\frac{c_{\beta\alpha}^2}{m_{h_0}^2}-\frac{1}{m_{A_0}^2}\bigg)\,,\\
C_{2}^{\prime}&=-\frac{1}{2}\big(\varepsilon_{23}^{d}\big)^2\bigg(\frac{s_{\beta\alpha}^{2}}{m_{H_0}^2}+\frac{c_{\beta\alpha}^2}{m_{h_0}^2}-\frac{1}{m_{A_0}^2}\bigg)\,,\\
C_{4}&=-\varepsilon_{23}^{d}\varepsilon_{32}^{d*}\bigg(\frac{s_{\beta\alpha}^{2}}{m_{H_0}^2}+\frac{c_{\beta\alpha}^2}{m_{h_0}^2}+\frac{1}{m_{A_0}^2}\bigg)\, .
\end{split}
\end{align}

Like in the case for $b\to s\ell^{+}\ell^{-}$, we only calculate a loop effect in the case of a vanishing tree-level contribution, i.e. for $\varepsilon^d_{23,32}=0$. In agreement with Ref.~\cite{Crivellin:2013wna} we find for the pure $H^+$ boxes
\begin{align}
\begin{split}
C_{1}&=-\frac{\big(V_{k2}^{*}\varepsilon_{kj}^{u}\varepsilon_{lj}^{u*}V_{l3}\big)\big(V_{m2}^{*}\varepsilon_{mi}^{u}\varepsilon_{ni}^{u*}V_{n3}\big)}{32\pi^2 m_{H^+}^2}I_{8}(z_j,z_i)\,,\\
C_{1}^{\prime}&=-\frac{\big(\varepsilon^{d*}_{22}\varepsilon^d_{33}V^{*}_{i 2}V_{i 3}\big)\big(\varepsilon^{d*}_{22}\varepsilon^d_{33}V^{*}_{j 2}V_{j 3}\big)}{32\pi^2 m_{H^+}^2} I_{9}(z_i,z_j)\,,\\
C_{2}&=-\frac{\big(\varepsilon_{22}^{d*}V_{j2}^{*}\varepsilon_{lj}^{u*}V_{l3}\big)\big(\varepsilon_{22}^{d*}V_{i2}^{*}\varepsilon_{ni}^{u*}V_{n3}\big)}{8\pi^2}\frac{\sqrt{z_i} \sqrt{z_j}}{m_{H^+}^2}I_{10}(z_i,z_j)\,,\\
C_{2}^{\prime}&=-\frac{\big(V^{*}_{n 2}\varepsilon^u_{n i}V_{i 3}\varepsilon^d_{33}\big)\big(V^{*}_{l 2}\varepsilon^u_{l j}V_{j 3}\varepsilon^d_{33}\big)}{8\pi^2}\frac{\sqrt{z_i} \sqrt{z_j}}{m_{H^+}^2}I_{10}(z_i,z_j)\,,\\
C_{4}&=-\frac{\big(\varepsilon_{22}^{d*}V_{j2}^{*}\varepsilon_{lj}^{u*}V_{l3}\big)\big(V_{m2}^{*}\varepsilon_{mi}^{u}V_{i3}\varepsilon_{33}^{d}\big)}{4\pi^2}\frac{\sqrt{z_i} \sqrt{z_j}}{m_{H^+}^2}I_{10}(z_i,z_j)\,,\\
C_{5}&=\frac{\big(\varepsilon_{22}^{d*}V_{j2}^{*}V_{j3}\varepsilon_{33}^{d}\big)\big(V_{m2}^{*}\varepsilon_{mk}^{u}\varepsilon_{nk}^{u*}V_{n3}\big)}{8\pi^2 m_{H^+}^2}\left(I_{8}(z_j,z_k)+I_{1}(z_j)\right)\,,
\end{split}
\end{align}
and for the $W^+$-$H^+$ boxes
\begin{align}
\begin{split}
C_{1}&=\frac{g_{2}^{2}}{64\pi^2}\frac{\sqrt{z_j}\sqrt{z_k}}{m_W^2}\big(V_{j2}^{*}\varepsilon_{ij}^{u*}V_{i3}\big)\big(V_{l2}^{*}\varepsilon_{lk}^{u}V_{k3}\big)I_{11}(y,z_k,z_j)\,,\\
C_4&=-\frac{ g_2^2\big(\varepsilon^{d *}_{2 2} \varepsilon^d_{3 3} V^{*}_{k 2} V_{k 3} V^{*}_{j 2} V_{j 3}\big)}{16\pi^2 m_W^2}I_{12}(z_j,z_k)\,.
\end{split}
\end{align}
The corresponding diagrams are shown in Fig. \ref{fig:Bs-mixing}.
The loop functions are given in the appendix and once more we did not distinguish between the cases of light and heavy quarks, {since the contribution of the light quarks trivially follows by taking the convergent limit $z_i\to 0$.}
\medskip

Phenomenologically, we only need to consider the contributions to $C_1$, since the other Wilson coefficients are proportional to $\varepsilon^d_{ij}$ which we will assume to be small. The constraints on NP crucially depend on the hadronic matrix elements calculated in lattice QCD. While Ref.~\cite{DiLuzio:2017fdq} finds a preference for destructive interference with the SM, Ref.~\cite{Bona:2007vi} finds a preference for constructive interference. We will therefore use the ratio $C_1^{\rm NP}/C_1^{\rm SM}$, where all hadronic uncertainties drop out. We assume a conservative bound of $\pm 30\%$.
\medskip

\begin{boldmath}
\subsection{$\ell\to\ell^{\prime} \gamma$ and $a_{\ell}$}
\end{boldmath}

Since it is important for our phenomenological analysis, we generalize the formula of Ref.~\cite{Crivellin:2013wna} to include right-handed neutrinos. Following the conventions of Ref. \cite{Crivellin:2018qmi} we define
\begin{align}
\mathcal{H}_{\rm eff}= c_{R}^{\ell_{F}\ell_{I}} \bar{\ell}_{F}\sigma_{\mu\nu}P_{R}\ell_{I}F^{\mu\nu} +{\rm h.c.}\,,
\end{align}
with
\begin{align}
a_{\ell_{I}}=-\frac{4m_{\ell_I}}{e}\Re\left[c_{R}^{\ell_{I}\ell_{I}}\right]\,,
\end{align}
and
\begin{align}
{\rm Br}\left[\ell_{I}\to\ell_{F}\gamma\right]=\frac{m_{\ell_I}^3}{4\pi}\tau_{\ell_I}\left(\big|c_{R}^{FI}\big|^2+\big|c_{R}^{IF}\big|^2\right)\,.
\end{align}
For the loop diagrams with charged Higgses we obtain
\begin{align}
\begin{split}
c_{R}^{\ell_F \ell_I}&=-\frac{e m_{\ell_I}\big(U_{Fk}\varepsilon_{kj}^{\nu}\varepsilon_{nj}^{\nu *}U_{In}^{*}\big)}{192\pi^2 m_{H^+}^2}\bigg[\frac{2x_{j}^2+5x_{j}-1}{(1-x_{j})^3}+\frac{6x_{j}^2\log(x_{j})}{(1-x_{j})^4}\bigg]+\frac{e m_{\ell_F}\varepsilon_{kF}^{\ell*}\varepsilon_{kI}^{\ell}}{192\pi^2 m_{H^+}^2}\,,\\
c_{L}^{\ell_F \ell_I}&=-\frac{e m_{\ell_F}\big(U_{Fk}\varepsilon_{kj}^{\nu}\varepsilon_{nj}^{\nu *}U_{In}^{*}\big)}{192\pi^2 m_{H^+}^2}\bigg[\frac{2x_{j}^2+5x_{j}-1}{(1-x_{j})^3}+\frac{6x_{j}^2\log(x_{j})}{(1-x_{j})^4}\bigg]+\frac{e m_{\ell_I}\varepsilon_{kF}^{\ell*}\varepsilon_{kI}^{\ell}}{192\pi^2 m_{H^+}^2}\,,
\end{split}
\end{align}
where we set the left-handed neutrino mass to zero. The neutral Higgs bosons give
\begin{align}
\begin{split}
c_{R}^{\ell_{F}\ell_{I}}&=\!\!\!\!\!\!\!\sum_{H=\{H_0,h_0,A_0\}}\!\!\!\!\!\!\!\!-\frac{e\left(m_{\ell_F}\Gamma_{jF}^{H*} \Gamma_{jI}^{H}+m_{\ell_I}\Gamma_{jF}^{H*}\Gamma_{jI}^{H}\right)}{192\pi^2 m_{H}^2} \!+\!\frac{e\, m_{\ell_{j}}\Gamma_{Fj}^{H}\Gamma_{jI}^{H}}{64\pi^2m_H^2} \Bigg(\!3+2\log\!\bigg(\frac{m_{\ell_j}^2}{m_{H}^2}\bigg)\!\Bigg)
\end{split}
\end{align}
with
\begin{align}
\Gamma_{FI}^{H_0}&=c_{\beta\alpha}\frac{m_{\ell_{F}}}{v}\delta_{FI}-s_{\beta\alpha}\varepsilon^{\ell}_{FI}\,, &
\Gamma_{FI}^{h_0}&=s_{\beta\alpha}\frac{m_{\ell_F}}{v}\delta_{FI}+c_{\beta\alpha}\varepsilon^{\ell}_{FI}\,, &
\Gamma_{FI}^{A_{0}}&=i\varepsilon_{FI}^{\ell}\,.
\end{align}
{Also here, we included the hard matching contribution together with the soft contribution from the effective theory in the formula since we do not aim at calculating QED corrections~\cite{Crivellin:2017rmk}.} For our purposes we require only the lepton flavour violating decay $\tau\to\mu\gamma$ whose experimental upper limit is given by $\mathrm{Br}\left[\tau\to\mu\gamma\right]<4.4\cdot 10^{-8}$ \cite{Aubert:2009ag,Hayasaka:2007vc}. 
\medskip

\begin{boldmath}
\subsection{$h\to\tau\mu$}
\end{boldmath}

Here, we find for the decay width
\begin{align}
\Gamma \left[h\to \tau\mu\right] \simeq \frac{3 c_{\beta\alpha}^2 m_h}{8\pi}  \Big(|\varepsilon_{23}^\ell|^2+|\varepsilon_{32}^\ell|^2\Big) \left(1-\frac{m_\tau^2}{m_h^2}\right)^2\,,
\end{align}
with $\Gamma_\mathrm{SM}\simeq 4.1{\rm MeV}$. This has to be compared to the current experimental limit~\cite{Aad:2016blu,Khachatryan:2015kon} 
\begin{align}
{\rm Br} \left[h\to \tau\mu\right] \le 1.43\%\,.
\end{align}
Due to the suppressed SM decay width, $h\to \tau\mu$ will turn out to be surprisingly constraining.
\medskip

\begin{figure}[t]
	\begin{center}
		\begin{tabular}{c}
			\hspace{5mm}	\includegraphics[width=0.7\textwidth]{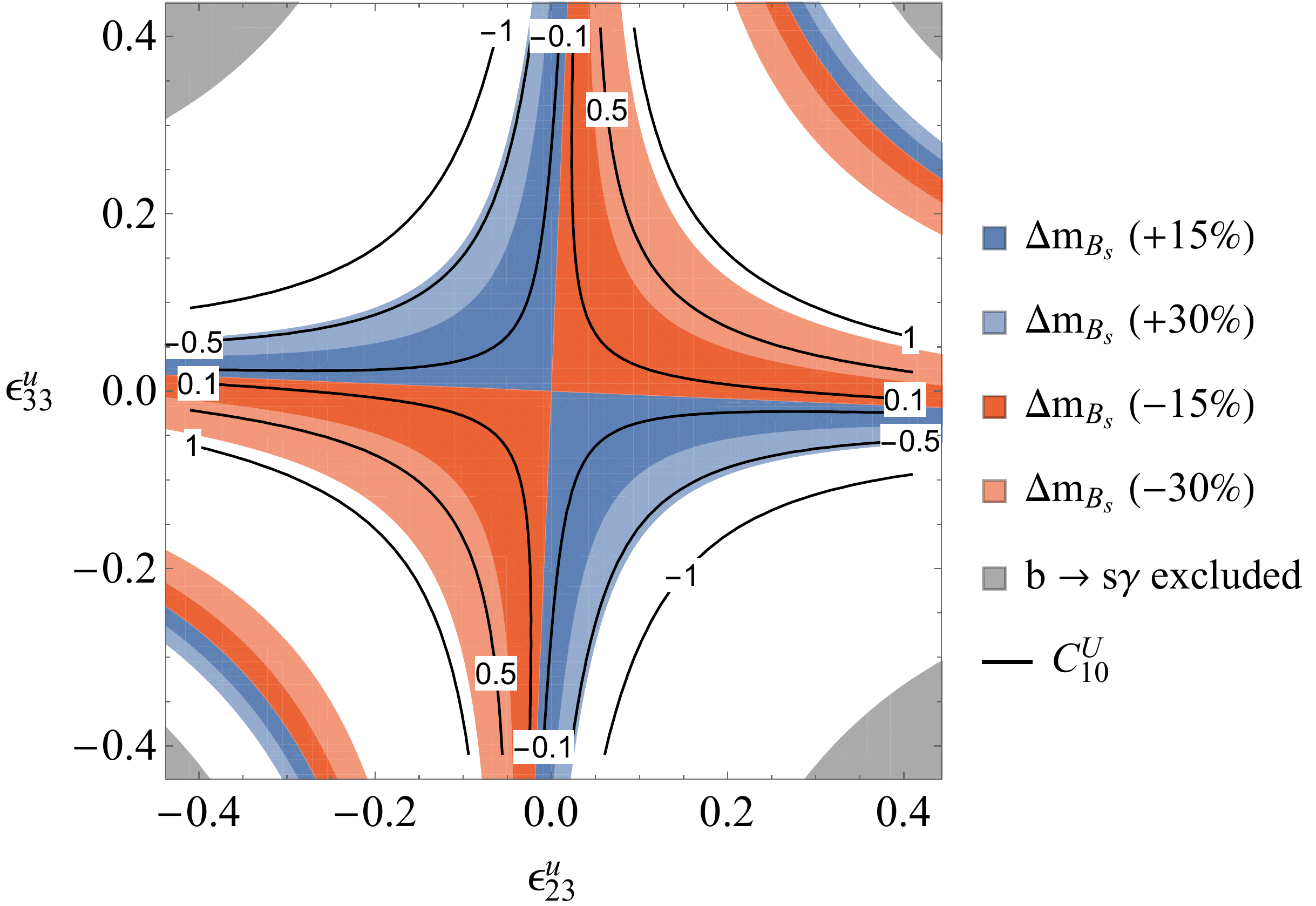}
		\end{tabular}
	\end{center}
	\caption{Effect in $B_s-\bar B_s$ mixing and $C_{10}^U$ in the $\varepsilon^{u}_{23}$-$\varepsilon^{u}_{32}$ plane for $M_{H^+}=400\mathrm{GeV}$ assuming all other couplings $\varepsilon=0$. Note that the relative effect in $C_{10}^U$ with respect to the one in $B_s-\bar B_s$ mixing is to a good approximation independent of the Higgs masses. The small allowed regions in the bottom-left (top-right) of the plot correspond to cancellations between boxes with two charged Higgses and mixed boxes with $W$ and $H^{\pm}$. \label{Plot1}}         
\end{figure}

\section{Phenomenological Analysis}
\label{phenomenology}

In our numerical analysis we want to focus on the possibility to explain the hints for NP in $b\to s\mu^+\mu^-$ transitions and $a_\mu$ within 2HDMs. Concerning $b\to s\mu^+\mu^-$ data, it is well-known from global fits that a sizeable contribution to the Wilson coefficient $C_9$ (and possibly also $C_{10}$) is required to explain the data. Additional substantial effects in $C_9^\prime$ and $C_{10}^\prime$ are possible. However, contributions to scalar operators must be suppressed due to the strong constraints from $B_s\to \mu^+\mu^-$ where they enter with an enhancement factor of $m_b^2/m_\mu^2$.
\medskip

\begin{figure}[t]
	\begin{center}
		\begin{tabular}{c}
			\hspace{5mm}	\includegraphics[width=0.7\textwidth]{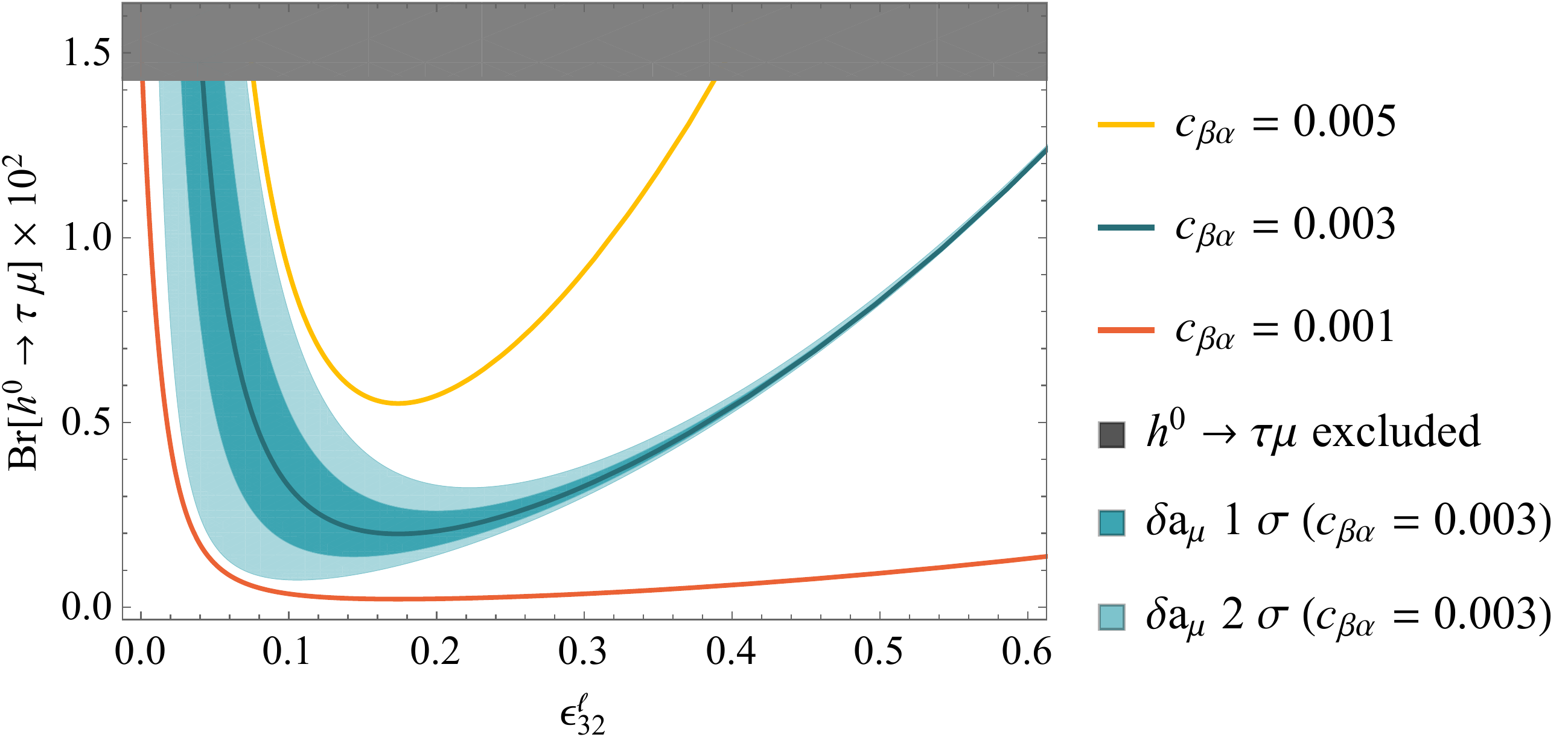}
		\end{tabular}
	\end{center}
	\caption{Prediction for the decay of the SM-like Higgs boson $h\to \tau\mu$ as a function of $\varepsilon^{\ell}_{32}$ under the assumption that $\varepsilon^{\ell}_{23}$ is chosen in such a way that the anomalous magnetic moment of the muon is explained. We used $M_{H^+}=400\mathrm{GeV}$, $M_{H_0}=250\mathrm{GeV}$ and $M_{A_0}=300\mathrm{GeV}$. For $c_{\beta\alpha}=0.003$ the whole $2\,\sigma$ region to explain $a_\mu$ is shown while for $c_{\beta\alpha}=0.001$ and $c_{\beta\alpha}=0.005$ only the predictions for the central value of $a_\mu$ are depicted.\label{Plot2}}         
\end{figure}

$C_9$ and $C_{10}$ can only be generated from $\gamma$ and $Z$ penguins (see Eqs. \eqref{gammaZ}-\eqref{light_quark_gamma_pen}) or from charged Higgs boxes (see \eq{HHboxes}). Interestingly, all contributions to $C_9$ and $C_{10}$ involve $\varepsilon^u_{ij}$ but not $\varepsilon^d_{ij}$ while the effect in $C_9^\prime$, $C_{10}^\prime$ only appears once $\varepsilon^d_{ii}$ is unequal to zero. Furthermore, scalar operators involve both $\varepsilon^d_{ii}$ and $\varepsilon^u_{ij}$. To accommodate the strong constraints on scalar operators we will assume that $\varepsilon^d_{ii}$ is negligibly small in the following. As stated above, an effect in $C_9$ is mandatory to explain the anomalies. However, the $Z$ penguin contribution to $C_9$ is suppressed by $(1-4s_W^2)$ and the off-shell photon effect is small due to the electromagnetic coupling. Hence, in the limit of $\varepsilon^\ell_{ij}=\varepsilon^\nu_{ij}=0$ we are left with a lepton flavour universal $C_{10}^U$ effect (following the conventions of Ref.~\cite{Alguero:2018nvb})  to a good approximation. This effect is also strongly correlated to (and therefore limited by) $B_s-\bar B_s$ mixing, as shown in Fig.~\ref{Plot1}. Note that this correlation is to a good approximation independent of the Higgs masses. The bound from $b\to s\gamma$ in this setup turns out to be in general weaker than the ones from $B_s-\bar B_s$ mixing.
\medskip

Therefore, we need in addition the charged Higgs boxes if we aim at a good fit to $b\to s\mu^+\mu^-$ data. {Here, $\varepsilon^\ell_{I2}$ generates $C_9^V=C_{10}^V$ effect in muons only, while $\varepsilon_{2I}^{\nu}$ gives $C_9^V=-C_{10}^V$}. Let us first consider the case with only $\varepsilon^\ell_{IJ}$ since these couplings are present also in the scenario without right-handed neutrinos. Since we aim at an explanation of $a_\mu$, we focus on the elements $\varepsilon^\ell_{23,32}$ which give an $m_\tau/m_\mu$ enhanced effect in this observable\footnote{Since it is a chirally enhanced effect, it has a free phase and can thus give a sizeable effect in the electric dipole moment of the muon~\cite{Crivellin:2018qmi}.}. For the numerical analysis we chose for definiteness $m_{A_0}=300\,$GeV and $m_{H_0}=250\,$GeV. Even though a detailed collider analysis is well beyond the scope of this article, note that the small values of $c_{\beta\alpha}$ are compatible with direct LHC searches~\cite{Sirunyan:2018koj}.  The effect in $a_\mu$ is directly correlated to $h\to\tau\mu$ which strongly constrains $c_{\beta\alpha}$ as shown in Fig.~\ref{Plot2}. The bounds from $h\to\tau\mu$ do not only depend on fewer parameters than $\tau\to\mu\gamma$ but are even much stronger for $\varepsilon_{22,33}^\ell=0$. Concerning $b\to s\ell^+\ell^-$, the impact with $\varepsilon^\ell_{23,32}\neq 0$ is small. Since the effect in $a_\mu$ is chirally enhanced, it significantly limits the product $\varepsilon^\ell_{23}\varepsilon^\ell_{32}$ rendering the deviation from $C_9^V=-C_{10}^V$ unimportant. 
\medskip

\begin{figure}[t]
	\begin{center}
		\begin{tabular}{c}
	\includegraphics[width=0.85\textwidth]{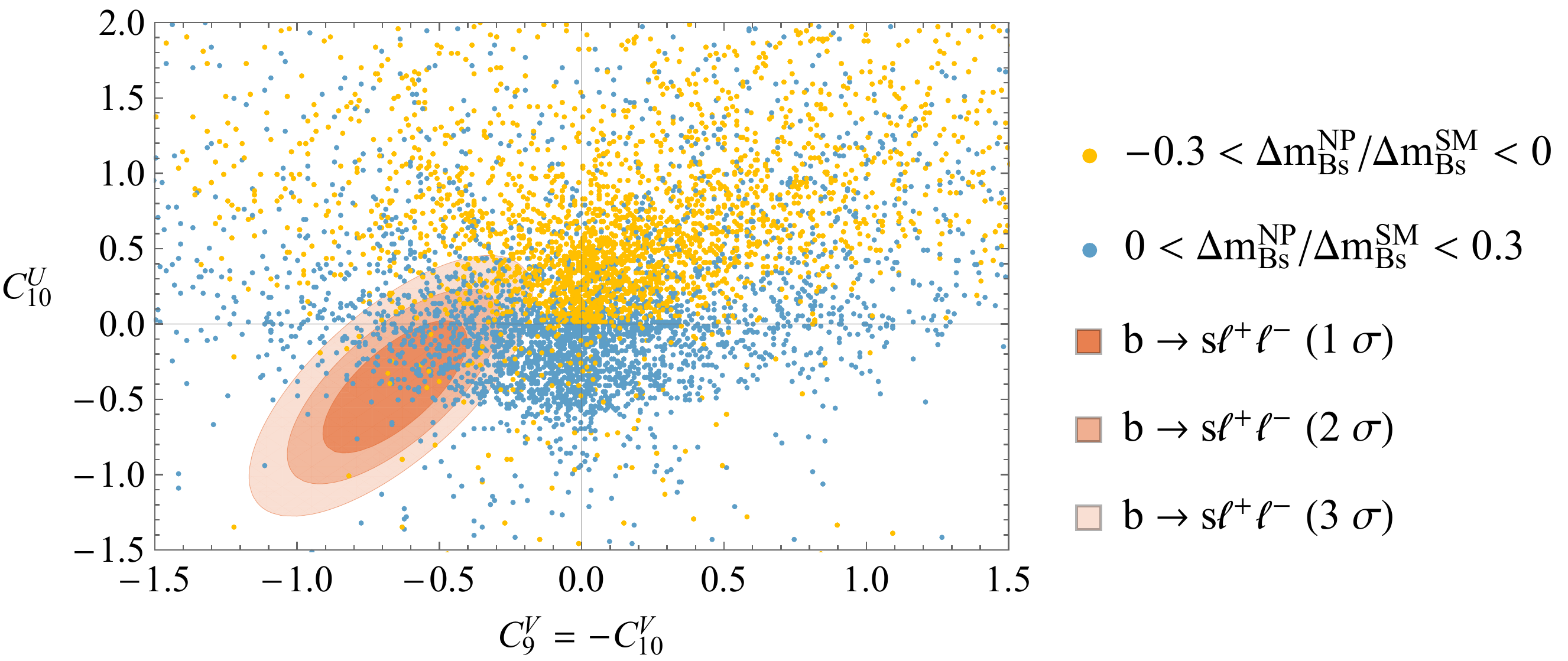}
		\end{tabular}
	\end{center}
	\caption{Scatter plot with $\varepsilon^u_{22,32,23,33}$ and $\varepsilon^\nu_{21,22,32,23,33}$ varied between $\pm 1.5$. Concerning the masses we scanned over are (in GeV) $m_{N_i}\in[100,1000]$, $m_{H^+}\in[100,500]$ and $\{m_{H_0},m_{A_0}\}\in[100,350]$. In total, we generated $10^6$ points. The red regions are preferred by $b\to s\ell^+\ell^-$ data according to updated fit of Ref.~\cite{Alguero:2019ptt} and includes the new LHCb~\cite{Aaij:2019wad} and Belle~\cite{BelleMoriond} measurement of $R(K)$ and $R(K^*)$, respectively. It is interesting to note that using the new fit significantly more points lie within the preferred regions. \label{Plot3}}         
 \end{figure}

In a next step, we allow for the presence of right-handed neutrinos and $\varepsilon_{ij}^{\nu}\neq 0$ where the $C_9^V=-C_{10}^V$ effect has to be added to $C_{10}^U$ from the $Z$ penguin. The result is shown in Fig.~\ref{Plot3} where we can see that it is difficult to find points which give a good fit to $b\to s\mu^+\mu^-$ data. 
While the effect of $\varepsilon_{IJ}^{\nu}\neq 0$ in $a_\mu$ is always destructive, i.e. it increases the discrepancy between theory and experiment, the effect is small since it is not enhanced by $m_\tau/m_\mu$. It is therefore possible to tackle $b\to s\mu^+\mu^-$ fixing $\varepsilon^\nu_{IJ}$ and $\delta a_\mu$ fixing $\varepsilon^\ell_{IJ}$ semi independently, while choosing the Higgs masses consistent with direct searches and taking into account the smallness of $c_{\beta\alpha}$, required by $h\to \tau\mu$. One can see that in order to be in agreement with $b\to s\ell^+\ell^-$ data, positive effects in $B_s-\bar B_s$ mixing are preferred.
\medskip

\section{Conclusions}
\label{conclusions}

In this article we studied $b\to s$ transitions in 2HDMs with generic Yukawa couplings (including right-handed neutrinos) with focus on $b\to s \mu^+\mu^-$ transitions and its possible correlations with $a_\mu$. We first recalled the tree-level effects in $b\to s$ observables which involve $\varepsilon^d_{23,32}$. If these elements are zero or negligibly small, loop effects involving $W$ bosons and charged Higgses can become numerically important. We calculated these leading one-loop corrections to $b\to s\ell^+\ell^-$, $b\to s\nu\bar{\nu}$ and $\Delta B=\Delta S=2$ transitions in a general $R_\xi$ gauge and confirmed their correctness finding gauge invariant results. Additionally, we discuss the treatment of self-energy contributions and renormalization in detail. In addition, we provided the formula for $\tau\to\mu\gamma$ and $a_\mu$ including the contributions from heavy (TeV scale) right-handed neutrinos.
\medskip

Concerning the phenomenology, we found that without right-handed neutrinos sizeable contributions to vector operators can only be generated via photon and $Z$ penguins. However, this does not allow for lepton flavour universality violation and the effect in $C_{10}^U$ with respect to $C_9^U$ is too big to give a good fit to data. Therefore, we included in a next step right-handed neutrinos which lead in general  to a lepton flavour universality violating $C_9^V=-C_{10}^V$ effect. This can provide an explanation of the anomalies especially with the recently updated $b\to s \ell^+\ell^-$ data.
\medskip

If we allow for Higgs to $\tau\mu$ couplings, we can explain the anomalous magnetic moment by a chirally enhanced $m_\tau/m_\mu$ effect. This leads at the same time to non-vanishing branching ratios $\tau\to\mu\gamma$ and $\tau\to3\mu$ which are however compatible with the experimental limits. The effect in $h\to\tau\mu$ is found to be dominant, i.e. most constraining. In case of an explanation of $a_\mu$, $h\to\tau\mu$ requires a close alignment in the Higgs sector, i.e. very small $c_{\beta\alpha}$. Furthermore, a small $C_9^V=+C_{10}^V$ effect is generated which does not significantly improve the goodness of the fit to data.
\medskip

2HDMs have a rich flavour phenomenology since they give effects in many classes of observables. As we showed in this article, these models are in principle capable to explain the discrepancies between the SM and experiment. Once one allows for a generic flavour structure and right-handed neutrinos, this provides a possible solution to the deviations in $b\to s\ell^+\ell^-$ transitions and $a_\mu$, even though some degree of finetuning is necessary. Furthermore, also the anomalies in $b\to c\tau\nu$ processes~\cite{Amhis:2016xyh} might be addressed by 2HDMs~\cite{Crivellin:2012ye,Crivellin:2013wna,Celis:2012dk,Ko:2012sv,Crivellin:2015hha,Dhargyal:2016eri,Chen:2017eby,Iguro:2017ysu,Martinez:2018ynq,Biswas:2018jun}. However, these solutions are under pressure from the measurement of the $B_c$ lifetime~\cite{Celis:2016azn,Alonso:2016oyd,Akeroyd:2017mhr,Blanke:2018yud} and LHC searches~\cite{Faroughy:2016osc}. Furthermore, also the $\epsilon^\prime/\epsilon$ anomaly (see e.g. Ref.~\cite{Aebischer:2018csl} for a review) could be explained~\cite{Chen:2018vog,Marzo:2019ldg}, leaving 2HDMs still as one of the most appealing NP  scenarios.

\section*{Acknowledgments}

The work of A.C. and D.M. is supported by an Ambizione Grant (PZ00P2\_154834) and a Professorship Grant (PP00P2\_176884) of the Swiss National Science Foundation. 
The work of C.W. is supported by the Swiss National Foundation under grant 200020\_175449/1. 
We are grateful to Bernat Capdevilla for providing us with the global fit used for Fig.~\ref{Plot3} and to Emanuele Bagnaschi for useful discussions. We thank Christoph Greub for collaboration in the early stages of the project and for useful comments on the manuscript.

\section*{Appendix}

We define the Higgs potential as
	\begin{align}
	\begin{split}
	\mathcal{V}(\Phi_1,\Phi_2)& = m_{11}^2\Phi_{1}^{\dagger}\Phi_{1} +m_{22}^2\Phi_{2}^{\dagger}\Phi_2-\big(m_{12}^2\Phi_{1}^{\dagger}\Phi_2+m_{12}^{2*}\Phi_2^{\dagger}\Phi_{1}\big)+\frac{\lambda_1}{2}\big(\Phi_{1}^{\dagger}\Phi_1\big)^2\\
	&~~+\frac{\lambda_2}{2}\big(\Phi_{2}^{\dagger}\Phi_{2}\big)^2+\lambda_3\big(\Phi_{1}^{\dagger}\Phi_{1}\big)\big(\Phi_{2}^{\dagger}\Phi_{2}\big) +\lambda_4 \big(\Phi_{1}^{\dagger}\Phi_{2}\big) \big(\Phi_{2}^{\dagger}\Phi_{1}\big)\\
	&~~+\frac{\lambda_5}{2}\Big(\big(\Phi_{1}^{\dagger}\Phi_{2}\big)^2+\big(\Phi_{2}^{\dagger}\Phi_{1}\big)^2\Big)\,.
	\end{split}
	\end{align}
Using the definition of Eq.~\eqref{Phidef} and transforming to the CP-even mass eigenstates according to Eq.~\eqref{baRot}, we express $m_{11}$, $m_{22}$, $m_{21}$, $\lambda_{1}$ and $\lambda_{4}$ in terms of the Higgs masses. Therefore, the remaining couplings are $\lambda_2$, $\lambda_{3}$ and $\lambda_{5}$. The triple Higgs couplings appearing in Eq.~\eqref{HiggsPenguRes} are then given by
\begin{align}
\begin{split}
\lambda_{h_0 H^+ H^-}&= v s_{\beta\alpha}\lambda_{3}\,,\\
\lambda_{H_0 H^+ H^-}&= v c_{\beta\alpha}\lambda_{3}\,.
\label{HiggsselfcouplingDef}
\end{split}
\end{align}
Note that with these conventions the expressions are as simple as possible and only $\lambda_3$ enters.

\subsection*{Loop Functions}
\label{App:loop_func}
The loop functions that we used throughout our article are defined as
\begin{align}
\begin{aligned}
f_{1}(b)&=\dfrac{{\left( {12{b}\left( {\log \left( {b} \right) - 1} \right) - 3b^2\left( {6\log \left(b\right) + 1} \right) + 8b^3 + 7} \right)}}{{{{\left( {1 - b} \right)}^4}}}\,,\\
f_{2}(b)&=\dfrac{{\left( {{4\log \left( b\right) + 3} - 2b\left( {3\log \left( b\right) + 4} \right) + 5b^2} \right)}}{{{{\left( {1 - b} \right)}^3}}}\,,\\
f_{3}(b)&=\dfrac{{\left( {3{b}\left( {2\log \left(b\right) + 1} \right) - 6b^2 + b^3 + 2} \right)}}{{{{\left( {1 - {b}} \right)}^4}}}\,,\\
f_{4}(b)&=\dfrac{{\left( { {2\log \left( b\right) + 3} - 4{b} + b^2} \right)}}{{{{\left( {1 - {b}} \right)}^3}}}\,,\\
f_{5}(b)&=\dfrac{{2\left( {12\log \left(b\right) + 19} \right) - 9{b}\left( {4\log \left( b \right) + 13} \right)}}{{{{\left( {1 - {b}} \right)}^4}}}+\dfrac{ 126b^2 + b^3\left( {18\log \left(b \right) - 47} \right)}{{{{\left( {1 - {b}} \right)}^4}}}\,,\\
I_0\left(b\right) &= \dfrac{1 - 3 b}{-1 + b} + \dfrac{2 b^2 \log\left(b\right)}{\left( b-1\right)^{2}}\,,\\
I_1\left(b\right) &= -\dfrac{1}{b-1} + \dfrac{\log\left(b\right) b}{\left( b-1\right)^{2}}\,,\\ 
I_2\left(b\right) &= \dfrac{\log\left(b\right) b}{1 - b}=(1-b)I_{1}(b)-1\,,\\ 
I_3\left(a, b\right) &= \dfrac{(7 a - b)b}{a - b}+\dfrac{2 b^2 \log\left(b\right) (2 a^2 - b^2 - 6 a + 3 b + 2 a b)}{\left(a - b\right)^{2} ( b-1)}- \dfrac{6 a^2 b \log\left(a\right)}{\left(a - b\right)^{2}}\,,\\
I_4\left(a, b\right) &= \dfrac{\sqrt{a^{3}} \sqrt{b} \log\left(a\right)}{(a-1) (a - b)} - \dfrac{\sqrt{a} \sqrt{b^{3}} \log\left(b\right)}{(b-1) (a - b)},\\ 
I_5\left(a, b\right) &= -1 + \dfrac{a^2 \log\left(a\right)}{( a-1) (a - b)} - \dfrac{b^2 \log\left(b\right)}{( b - 1) (a - b)},\\
I_6\left(b\right) &=- b + \dfrac{b^2 \log\left(b\right)}{ b-1 } = b(b-1)I_{1}(b),\\ 
I_7\left(b\right) &= \dfrac{b}{ b-1} - \dfrac{b^2 \log\left(b\right)}{\left( b-1\right)^{2}}=-bI_{1}(b)\,\\
I_{8}(a,b)&=\frac{-1}{(1-a)(1-b)}+\frac{b^2 \log(b)}{(1-b)^2(a-b)}+\frac{a^2 \log(a)}{(1-a)^2(b-a)}\,,\\
I_{9}(a,b)&=\frac{-ab}{(1-a)(1-b)}+\frac{ab\log(b)}{(1-b)^2(a-b)}+\frac{ab\log(a)}{(1-a)^2(b-a)}\,,\\
I_{10}(a,b)&=\frac{-1}{(1-a)(1-b)}+\frac{a\log(a)}{(1-a)^2(b-a)}+\frac{b\log(b)}{(1-b)^2(a-b)}\,,\\
I_{11}(a,b,c)&=\frac{-3a^2\log(a)}{(a-1)(a-b)(a-c)}+\frac{b(4a-b)\log(b)}{(b-1)(a-b)(b-c)}+\frac{c(4a-c)\log(c)}{(c-1)(a-c)(c-b)}\,,\\
I_{12}(a,b)&=\frac{ab\log(a)}{(1-a)(a-b)}-\frac{ab\log(b)}{(1-b)(a-b)}\,.
\end{aligned}
\end{align}
If the Higgs penguins contain a charm quark in the loop (whose mass we neglect), i.e. $z_2=0$, the loop functions simplify to
\begin{align}
\begin{aligned}
I_{0}(0)&=-1\,,\\
I_{1}(0)&=1\,,\\
I_{4}(b,0)&=I_{4}(0,b)=I_{4}(0,0)=0\,,\\
I_{5}(b,0)&=I_{5}(0,b)=-1+\frac{b\log(b)}{b-1}\,,\\
I_{5}(0,0)&=-1\,,\\
I_{7}(0)&=0\,,\\
I_{8}(0,x_j)&=I_{1}(x_j)\,,
\end{aligned}
\end{align}
and the corresponding Wilsons coefficients in \eq{HiggsPenguRes} become
\begin{align}
\begin{split}
C_{S(HH)}^{IJ}=&\dfrac{-y\varepsilon^{d *}_{2 2}L_{+}^{IJ}}{2 g_2^4 s_W^2 V^{*}_{t s} V_{t b}}\Big[
     4 t_2 \big(
       \varepsilon^d_{3 3} V^{*}_{k 2} \varepsilon^{u}_{k 2} V_{2 3}
      -\varepsilon^{d *}_{3 3} V^{*}_{2 2} \varepsilon^{u *}_{n 2} V_{n 3}
  \big)+ V^{*}_{k 2} \varepsilon^{u}_{k 2} \varepsilon^{u *}_{n 2} V_{n 3}
\\
&     -2 \log\left(\dfrac{\mu^2}{m_{H^+}^2}\right) \Big(
      2 \left(\varepsilon^d_{3 3} V^{*}_{k 2} \varepsilon^{u}_{k 2} V_{2 3}
      -\varepsilon^{d *}_{3 3} V^{*}_{2 2} \varepsilon^{u *}_{n 2} V_{n 3}\right) t_2\\
&       +  V_{n 3} (2 V^{*}_{2 2} \varepsilon^{u*}_{n 2} \varepsilon^{u}_{2 2} 
        + 2 V^{*}_{2 2} \varepsilon^{u*}_{n 3} \varepsilon^{u}_{2 3} 
        + 2 V^{*}_{3 2} \varepsilon^{u*}_{n 2} \varepsilon^{u}_{3 2} 
        - V^{*}_{k 2} \varepsilon^{u*}_{n 2} \varepsilon^{u}_{k 2})
  \Big)\\
&          -4\left( V^{*}_{2 2} \varepsilon^u_{2 2} \varepsilon^{u *}_{n 2} V_{n 3}
-I_5\left(z_3, 0\right)\left( V^{*}_{2 2} \varepsilon^u_{2 3} \varepsilon^{u *}_{n 3} V_{n 3}
+ V^{*}_{3 2} \varepsilon^u_{3 2} \varepsilon^{u *}_{n 2} V_{n 3}\right)\right)
\Big]
\end{split}
\end{align}
\begin{align}
\begin{split}
C_{S(HH)}^{\prime IJ}\!=&\dfrac{y L_{-}^{IJ}}{g_2^4 s_W^2 V^{*}_{t s} V_{t b}}\! \!\Big[
     \!-\!2  t_2  \big(\!
       \left(\varepsilon^d_{3 3}\right)^{\!2}  V^{*}_{k 2} \varepsilon^{u}_{k 2} V_{2 3}
      \!-\!\varepsilon^{d *}_{2 2} \varepsilon^d_{2 2} V^{*}_{2 2} \varepsilon^{u *}_{n 2} V_{n 3}
  \big)\\
&    +\!2 \log\left(\dfrac{\mu^2}{m_{H^+}^2}\!\right)\!\Big(\!\!
       -\!\varepsilon^d_{3 3} V^{*}_{k 2} \varepsilon^{u}_{k 2} \varepsilon^{u *}_{2 2} V_{2 3}
       -\!\varepsilon^d_{3 3} V^{*}_{k 2} \varepsilon^{u}_{k 2} \varepsilon^{u *}_{3 2} V_{3 3}
       -\!\varepsilon^d_{3 3} V^{*}_{k 2} \varepsilon^{u}_{k 3} \varepsilon^{u *}_{2 3} V_{2 3}
      \!\\
&+\!\left(\!\left(\varepsilon^d_{3 3}\right)^{\!2}\! V^{*}_{k 2} \varepsilon^{u}_{k 2} V_{2 3}\!-\!\varepsilon^{d *}_{2 2} \varepsilon^d_{2 2} V^{*}_{2 2} \varepsilon^{u *}_{n 2} V_{n 3}\!\right)\! t_2
  \Big)
      -2 \varepsilon^d_{3 3}V^{*}_{k 2} \varepsilon^{u}_{k 2} \varepsilon^{u *}_{2 2} V_{2 3}\\
&      + \varepsilon^d_{3 3}\big(
       - \varepsilon^{d *}_{2 2} \varepsilon^d_{2 2} V^{*}_{2 2} V_{2 3}
      +2 I_5\left(z_3, 0\right) V^{*}_{k 2} \left(\varepsilon^{u}_{k 3} \varepsilon^{u *}_{2 3} V_{2 3}+
         \varepsilon^{u}_{k 2} \varepsilon^{u *}_{3 2} V_{3 3}\right)
  \big)
\Big]
\end{split}
\end{align}
\newpage
\bibliographystyle{JHEP}
\bibliography{BIB} 

\end{document}